\documentclass[final,3p,times]{elsarticle}

\usepackage{amsmath}
\usepackage{amssymb}
\usepackage[usenames,dvipsnames]{color}

\biboptions{sort&compress}

\journal{Physica A}

\begin{document}

%%%%%%%%%%%%%%%% FRONT MATTER %%%%%%%%%%%%%%%%%%%%%%%%

\begin{frontmatter}

\title{Unidirectional Random Growth with Resetting}

\author{T.~S.~Bir{\'o} \corref{c1}}
\ead{Biro.Tamas@wigner.mta.hu}
\cortext[c1]{Corresponding author}
\address{Theory Department, H.A.S. Wigner RCP, Budapest, Hungary}

%%%%%%%%%%%%%%%%%%%%%%%%%%%%%%%%%%%

\author{Z.~N\'eda \corref{c2} }
\ead{zneda@phys.ubbcluj.ro}
\cortext[c2]{Contributing author}
\address{Department of Physics, Babe\c{s}-Bolyai University, Cluj-Napoca, Romania}

%%%%%%%%%%%%%%%%%%%%% DOCUMENT BODY %%%%%%%%%%%%%%%%%%%%%

\begin{abstract}
%\textcolor{red}{Break it to shorter sentences.}

%Unidirectional processes cannot satisfy the detailed balance condition.
We review and classify stochastic processes without detailed balance condition.
We obtain stationary distributions and investigate their stability in terms of
generalized entropic distances beyond the Kullback-Leibler formula.
A simple stochastic model with local growth rates and direct resetting to the ground state
is investigated and applied to various networks, scientific citations and Facebook popularity,
hadronic yields in high energy particle reactions, income and wealth distributions,
biodiversity and settlement size distributions.
\end{abstract}


\end{frontmatter}

%\maketitle

\newcommand{\be}{\begin{equation}}
\newcommand{\ee}[1]{\label{#1} \end{equation}}

\newcommand{\ba}{\begin{eqnarray}}
\newcommand{\ea}[1]{\label{#1} \end{eqnarray}}
\newcommand{\nl}{\nonumber \\}
\newcommand{\pd}[2]{ \frac{\partial {#1}}{\partial {#2}} \,}
\newcommand{\pt}[2]{ \frac{{\rm d} {#1}}{{\rm d} {#2}} \,}
\newcommand{\exv}[1]{ \left\langle {#1} \right\rangle }
\newcommand{\eon}[1]{ \mathrm{e}^{ {#1} } }

\renewcommand{\Im}{\, \mathfrak{Im}\, }
\renewcommand{\Re}{\, \mathfrak{Re}\, }

\newcommand{\sumi}[1]{ \sum_{{#1} = 0}^{\infty}\limits }
\newcommand{\inti}{ \int_0^{\infty}\limits\! }
\newcommand{\intii}{ \int_{-\infty}^{+\infty}\limits\! }

\newcommand{\re}[1]{ eq.(\ref{#1}) }

%%%%%%%%%%%%%% 

\section{Introduction}

%\textcolor{Blue}{
%\begin{itemize}
%\item complex systems, stochastic dynamics
%\item stationary probability distributions
%\item non-exhaustive review of applied methods and problems
%\item convergence and entropic distance derived from master equations
%\item we generalize entropic distance and entropy formulas
%\end{itemize}
%}

The challenge for physicists in taming complexity is to identify clear and simple models with
possibly few ingredients and a great and rich reign of applicability.
We have in mind achievements like the Ising model of magnetism \cite{ISING}, the Erd\H{o}s--R\'enyi 
random graph model \cite{ERDOSRENYI,GILBERT}, or the Landau $\Phi^4$--theory for second order
phase transitions \cite{LANDAU}. These models have their beauty and usefulness not only in describing
particular physical phenomena, but also in allowing for the gain of new fundamental insights.
The Ising model led us to investigate critical behavior, the Erd\H{o}s--R\'enyi graph
to abundant research on path length and other optimization problems on random networks,
and the Landau theory opened the door to study in a unified framework 
all types of phase transitions.

Master equations describing stochastic processes belong to a similar model class with a wide range
of applicability to complex systems \cite{MASTER}.
The question of stability of stationary solutions to such equations is recently
connected to fundamental questions about the notion and correct mathematical treatment of
entropy and entropic divergence.
Most approaches using master equations contain equally growth and loss terms, and very often 
detailed balance condition is tacitly assumed. In several cases the stationary solution is
presented, but the convergence rate to it is not elaborated.

In this paper we attempt an in-depth study of a particular class of stochastic processes:
where the growth process dominates and only a very special transition to a ground state
is allowed. This restriction immediately breaks the detailed balance condition.
On the other hand such processes belong to the sample space reduction types in which
a growing interest can be documented recently \cite{ThurnerSSR}. Despite of its simplicity this approach
offers a rich variety of complex behavior with corresponding probability distribution
functions (PDF-s). In full agreement with the physicist's philosophy for handling complex
systems the particular model we present here is based on only two dynamical ingredients:
a growth rate and a reset rate.

Going beyond this special class of master equations,
in a general framework for nonlinear stochastic models we prove the reduction of entropic
divergence. For a power-law dependence on the probabilities we find a formula
generalizing the well-known Kullback--Leibler result \cite{KULLBACKLEIBLER}. 
We relate the entropy -- probability relation to the entropic divergence expression by using 
the uniform distribution
as a reference. While in the classical logarithmic formula from this comparison
the Boltzmann--Gibbs--Shannon entropy arises, in a more general case the entropic divergence
cannot simply be treated as relative entropy (difference of entropies).
In particular for a power-law nonlinearity in the master equation we arrive at an entropic
divergence formula which is proportional, but not equal to, a difference of Tsallis
$q$-entropies. For the original presentation of Tsallis and earlier suggestions for
generalized information measures see 
Refs. \cite{TSALLISENTROPY,TSALLISPRE1,TSALLISPRE2,ENTSUM,RECENTINFO}.
For nonlinearly modified Fokker--Planck equation, leading to power-law tailed $q$-exponential
and $q$-Gaussian distributions, also in connection with non-Boltzmannian entropy
formulas we refer to \cite{NLFOKKER,GENFOKKER,HTHEOREMNLFP,qGAUSSless,qGAUSSmore}.
Entropy production according to the second law of thermodynamics also has been studied
in this framework \cite{CARNOTq,PARTICLES,SECONDLAWq}.

In the present paper common distributions are reproduced in the framework of the growth and reset model
focusing on the continuous limit. A special emphasis is put on the growth rate
with linear preference (Matthew principle \cite{Matthew}). The reset can also be due to an exponential
dilution of the sample space, not necessarily describing direct transitions to the
ground state. This represents a further generalization for the applicability of
this model framework.

We point out that not only the direct problem of obtaining stationary PDF-s from
known transition rates can be handled, but also the reverse problem of finding
the correct transition rate formula from the known PDF. This is important for
many practical applications. Furthermore in some applications, like networks,
one can measure both transition rates and PDF-s, offering a testground for our
model master equation.

Following all these theoretical elaborations we present a number of real-world
applications. We mostly select those which are familiar and fashionable among
the statistical physics community: networks, citation statistics, multiplicity distributions
in high energy particle and heavy-ion collisions, income and wealth distributions,
biodiversity and city size frequencies. Applications of different $q$-nonlinear
dynamical models, e.g. description of vortex dynamics in type II superconductors
are given in Refs. \cite{SUPERIIq,SUPERIIvortex}, and will not be discussed in this review.

Besides reading all sections in the order given, one may have a first look directly
starting with the applications. However, for readers interested in fundamental problems
in thermodynamics and statistical physics we recommend to dive into 
the details in the leading sections.

Finally we have to apologize for not presenting a classical review with detailed
reference to all important achievements on the field of complex systems and the entropy
generalization. Our more modest intention here is to promote an elaboration on the
simple idea of a view of local growth balanced with nonlocal resetting transitions.

%%%%%%%%%%%%% MASTER EQ %%%%%%%%%%%%

\section{Evolution Master Equations}

%\textcolor{Blue}{
%\begin{itemize}
%\item Linear Master Equation
%\item Transition rates: examples
%\item Mixed short and long jumps
%\item Missing detailed balance
%\end{itemize}
%}

In complex systems frequently non-exponential, mostly  power-law tailed distributions emerge as stationary ones. 
It is particularly intriguing the case when such stationary distributions result from unbalanced, unidirectional
processes \cite{BiroNedaPRE2017}. 
This behavior is in contrast to the classical diffusion, where sizable transition probabilities 
are for choosing opposite directions.
In this paper we mainly deal with processes when the state variable, briefly noted
by a natural number $n$, or by its continuous counterpart, $x$, evolves only in one
direction, limiting ourselves to Markovian processes.
Such unbalanced growth processes alone do not lead to nontrivial stationary distributions,
just to a simple inverse proportionality with the local transition rates. 
Therefore we add a special nonlocal transition possibility from any state $n$ back to the $n=0$ ground state. 

All transition rates, $w_{nm}$,
starting from a state $m$ and ending in a state $n$ are coefficients to the respective probabilities
in the evolution master equation. To start simply, we assume a linear dependence and consider
\be
 \dot{P}_n \: = \: \sum_m\limits \left[w_{nm}P_m - w_{mn} P_n \right].
\ee{STARTLINEARDYN}
Here the antisymmetric structure of the right hand side ensures that $\sum_n\limits \dot{P}_n = 0$, so this
evolution equation conserves the normalization $\sum_n\limits P_n = 1$.
We emphasize at this point that we will later also consider nonlinear dependence on the $P_n$ probabilities.

\begin{figure}
\begin{center}
 \includegraphics[width=0.75\linewidth]{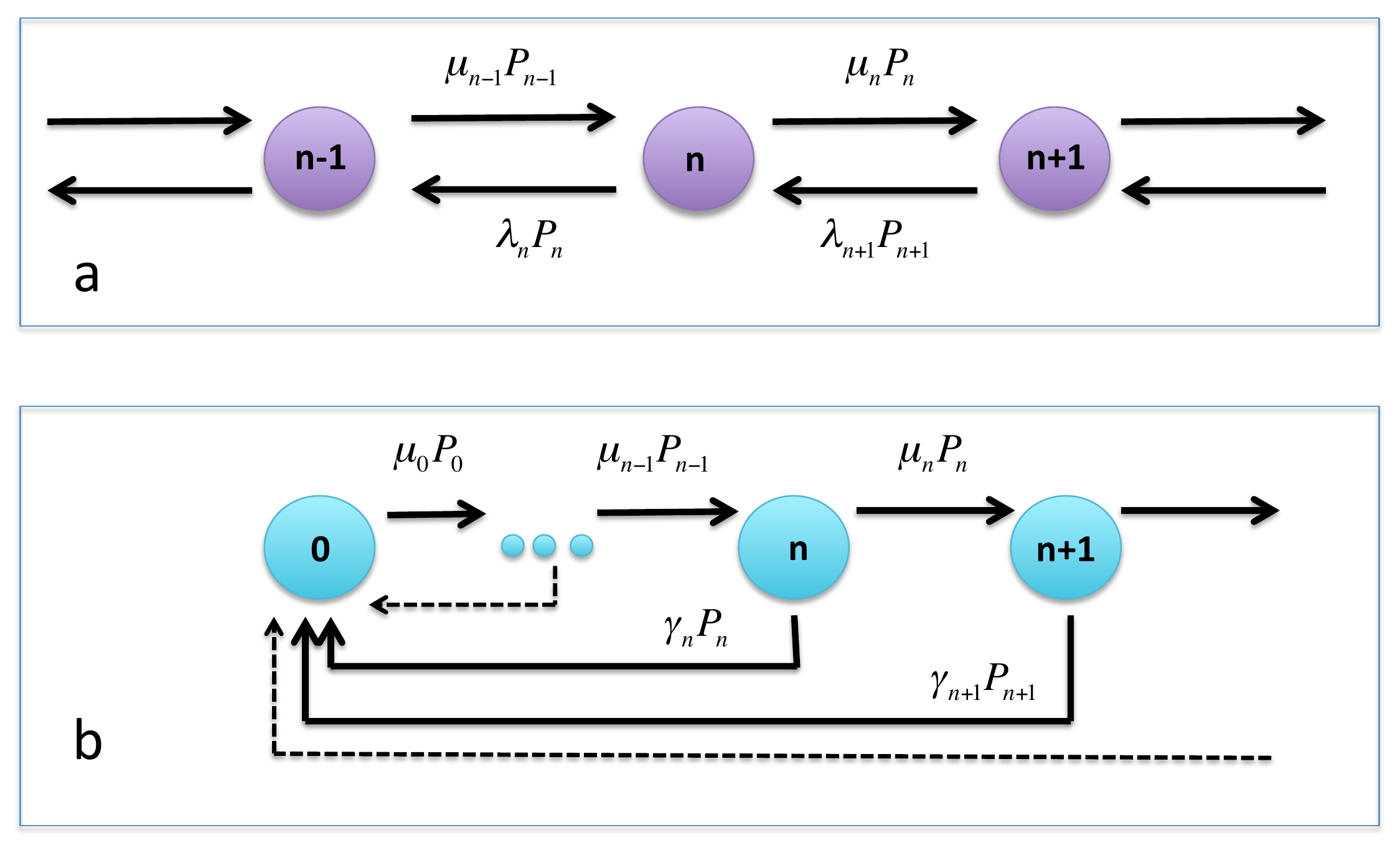}

\end{center}
\caption{ \label{pic2a}
 Markov chain for the symmetric local process with general state-dependent transition rates
(top, Figure a) and for the unidirectional growth augmented by rare resets (bottom, Figure b).}
\end{figure}

A well-known example is the classical diffusion with a drift \cite{CLASSICDIFFUSION,MATHDIFFUSION,DIFFSURVEY}. 
In this case $n$ can be both higher and lower than $m$, but only
with a single unit. The transition rates are thus local,
\be
 w_{nm} \: = \: \mu_m \delta_{n,m+1} + \lambda_m \delta_{n,m-1},
\ee{STARTDIFFUSION}
leading to the evolution equation (sketched in Fig.\ref{pic2a}, top):
\be
 \dot{P}_n \: = \: \mu_{n-1}P_{n-1} + \lambda_{n+1}P_{n+1} - \mu_nP_n - \lambda_n P_n.
\ee{STARTDIFFUEVEQ}
This equation is a discrete model of the one-dimensional diffusion, with position dependent drift
and diffusion coefficients. In the continuous model we define the probability density function
${\cal P}(n \Delta x, t) = P_n/ \Delta x$. The normalization is given by $\sum_n\limits P_n = \int\!{\cal P}(x,t) dx = 1$.
The drift coefficient is defined by $v(n\Delta x) = (\lambda_n-\mu_n) \Delta x$, the diffusion coefficient by
$D(n \Delta x) = (\lambda_n+\mu_n) \Delta x^2/2$. This leads to the Fokker-Planck equation:
\be
 \pd{{\cal P}}{t} \: = \: \pd{}{x} (v {\cal P}) + \pd{^2}{x^2} ( D{\cal P}).
\ee{CONTIDIFFUSEQ}
%The corresponding discrete state, continuous time Markov chain is depicted in Fig.\ref{pic2a}.
Such systems are well-studied \cite{FOKKERPLANCK} 
and several theorems are known about their stationary distributions and stability,
describing among others the evolution towards the stationary state.

We would like to deal here, however, with extremely different processes, typically with those when the
transitions between elementary states are unidirectional, increasing $n$ by one unit. This local
growth rate is supplemented by a nonlocal transition, directly resetting any state
to ground state ($n=0$):
\be
 w_{nm} \: = \: \mu_m \delta_{n,m+1} + \gamma_m \delta_{n,0}.
\ee{WUNIRESET}
We refer to such stochastic evolutions as ''unidirectional growth with resetting''. 
The above transition rates after summation over the starting states $m$ lead to
the following time evolution:
\be
 \dot{P}_n \: = \: \mu_{n-1} P_{n-1} + \delta_{n,0} \exv{\gamma} \,  - \, (\mu_n + \gamma_n) P_n.
\ee{PDOTUNIRESET}
In the above equation $\exv{\gamma}=\sum_j\limits \gamma_j P_j$ and only $n \ge 0$
is possible.  The Markov chain for such systems is depicted in Fig.\ref{pic2a} (bottom).
Further simplification may arise when one specifies the step-up rates, $\mu_n$, and the reset rates, $\gamma_n$.
The simplest choice with constant transition rates, although already powerful, does not describe the most
interesting phenomena. On the other hand state-dependent rates lead to several nontrivial distributions.
In particular we study growth rates linear in $n$ with constant reset rates, 
and discuss further possibilities, too. 
%Each fixing of the rates $\mu_n$ and $\gamma_n$ induces a different
%time evolution towards different stationary distributions.

%%%%%%%%%%%%%%%%%%%%%%%%%%%%%%%%%%%%%%%%%%%%

\section{Entropic Divergence Evolution}

%\textcolor{Red}{
%\begin{itemize}
%\item Proof without detailed balance
%\item Generalization to nonlinear master equations
%\item Example: Kullback-Leibler vs. q-log divergence
%\item Problem if initial and final state dependencies are both present
%\end{itemize}
%}

Entropy plays an important role in studies of convergence and stability of random processes.
The purpose of this section  is  to find the proper expression for entropy for
growth processes that do not fulfill the detailed balance condition. 
While studying the convergence of arbitrary initial distributions towards the stationary one,
we construct the entropic distance measure which has to shrink. 
We derive new generalized entropy formulas based on the entropic distance tailored on a given
class of dynamics.

\subsection{Decreasing entropic distances for processes without detailed balance}

First we have a look at linear master equations with transition rates $w_{nm}$
which are not satisfying the detailed balance condition,
\be
 \frac{w_{nm}}{w_{mn}} \: \ne \: \frac{Q_n}{Q_m}.
\ee{DETBALNM}
The stochastic dynamics is described by the following linear master equation
\be
 \dot{P}_n \: = \: \sum_m\limits \left[ w_{nm}\, P_m - w_{mn}\, P_n \right].
\ee{LINMASTER}
The stationary distribution by definition satisfies merely the following total balance
\be
 0 \: = \: \sum_m\limits \left[ w_{nm}\, Q_m - w_{mn}\, Q_n \right].
\ee{LINSTATIO}
This can be rearranged into the well known ''self averaging'' form:
\be
 Q_n \: = \: \frac{\sum_m\limits w_{nm} Q_m }{\sum_m\limits w_{mn}}.
\ee{QSELFAVER}
We would like to construct an entropic divergence formula, represented by
the real-valued functional, $\rho[P,Q]$,  whose shrinking describes the approach from any actual 
probability distribution $P_n(t)$ towards the stationary one, $Q_n$. The functional
$\rho[P,Q]$  has to respect the following general properties:
\begin{enumerate}
\item $\rho[P,Q] \ge 0$ for any pair of normalized distributions $P_n$ and $Q_n$.
\item From $\rho[P,Q]=0$ it uniquely follows that $P_n=Q_n$.
\item With $Q_n$ being the stationary distribution it  evolves according to $\dot{\rho}[P,Q] \le 0$.
\item $\dot{\rho}[P,Q]=0$ achieved only if $P_n=Q_n$.
\end{enumerate}
We note by passing that symmetry is not necessary.
We seek the entropic divergence in the special scaling trace form
\be
 \rho[P,Q] \: = \: \sum_n\limits Q_n \, \sigma(\xi_n),
\ee{EDIVFORM}
with $\xi_n=P_n/Q_n$ and $\sigma(\xi)$ a function whose properties we discuss later. 
This distance measure is not symmetric for the exchange of $P$ and $Q$, therefore
it cannot be used for extracting metric properties, like the triangle inequality in the space of possible distributions.
A symmetrization, however, can be achieved by using the sum
\be
 \rho[P,Q] + \rho[Q,P] \: = \: \sum_n\limits Q_n \, \left[\sigma(\xi_n) + \xi_n \sigma(1/\xi_n)  \right].
\ee{RHOSUM}
This is equivalent with using the core function
\be
 \kappa(\xi) \: \equiv \: \sigma(\xi) + \xi \, \sigma(1/\xi)
\ee{SYMSIGMA}
in each term in the sum.

\begin{figure}[h]
\center{\includegraphics[width=0.7\textwidth]{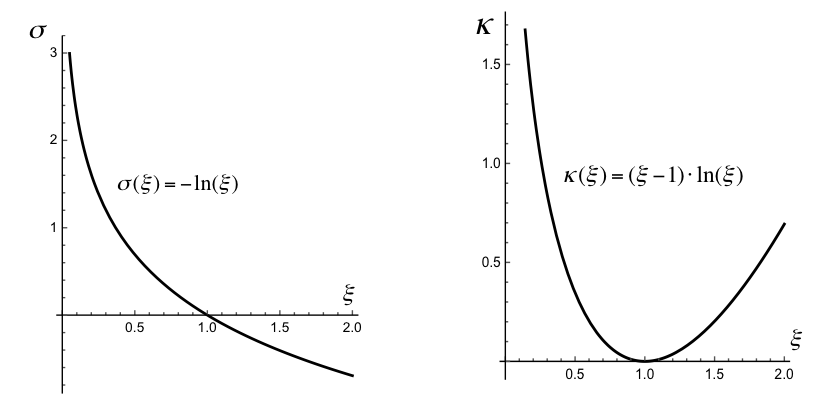}}

\caption{\label{FIG:COREFUNCTION}
 An example for the core functions $\sigma(\xi)=-\ln \xi$ and its $[P,Q]$ symmetrized counterpart 
 $\kappa(\xi) = (\xi-1)\ln\xi$.
}
\end{figure}

The property to have non-negative entropic divergence can be related to the curvature of the
core function $\sigma(\xi)$ by making use of the Jensen inequality:
\be
 \sum_n\limits Q_n \sigma(\xi_n) \: \ge \: \sigma\Big(\sum_n\limits Q_n\xi_n \Big) 
 \: = \: \sigma\Big( \sum_n\limits P_n \Big) \: = \: \sigma(1)
\ee{SIGJENSEN}
for $\sigma^{\prime\prime} > 0$. For satisfying properties 1 and 2 one uses core functions
with overall positive second derivative and $\sigma(1)=0$. From these conditions also follows
$\kappa(1)=0$ and $\kappa^{\prime\prime} > 0$.
Studying the first and second derivative of the symmetrized distance's core function,
$\kappa(\xi)$, we arrive at an interesting consequence. One easily obtains
$\kappa(1)=2\sigma(1)=0$, $\kappa^{\prime}(1)=0$ and $\kappa^{\prime\prime}>0$.
This means that $\xi=1$ is the minimum of $\kappa(\xi)$ with the value zero, therefore
for all ratios $\xi_n=P_n/Q_n$ we have a positive contribution to the symmetrized distance:
$\kappa(\xi) \ge 0$ (cf. Fig.\ref{FIG:COREFUNCTION}).

According to the trace form (\ref{EDIVFORM}) the time derivative of the entropic divergence 
from the stationary distribution, $Q_n$, is given as follows
\be
 \dot{\rho} \: = \: \sum_n\limits \sigma^{\prime}(\xi_n) \, \dot{P}_n 
 \: = \: \sum_{n,m}\limits \sigma^{\prime}(\xi_n) \, \big( w_{nm}Q_m\xi_m - w_{mn}Q_n\xi_n \big).
\ee{RHODOT}
%Here we replaced $P_m=Q_m\xi_m$ and $P_n=Q_n\xi_n$.
Applying the identity $\xi_m = \xi_n + (\xi_m-\xi_n)$ we get
\be
 \dot{\rho} \: = \: \sum_{n}\limits \xi_n \sigma^{\prime}(\xi_n) \sum_m \limits \left(w_{nm}Q_m-w_{mn}Q_n \right)
 + \sum_{n,m}\limits \sigma^{\prime}(\xi_n) \, (\xi_m-\xi_n) \, w_{nm}Q_m.
\ee{DIFFFORM}
Here the first sum contains a term which is zero due to the total balance condition eq.(\ref{LINSTATIO}).
The sign of the remaining terms in the double sum depends on the factor $\sigma^{\prime}(\xi_n)(\xi_m-\xi_n)$.
This is the first order term in the Taylor expansion. Indeed the remainder theorem for Taylor series in
the Lagrange form ensures that
\be
 \sigma(\xi_m) \: = \: \sigma(\xi_n) + \sigma^{\prime}(\xi_n) \, (\xi_m-\xi_n) + \frac{1}{2}
 \sigma^{\prime\prime}(\xi_{mn}) \, (\xi_m - \xi_n)^2
\ee{SIGTAYLOR}
with some $\xi_{mn}$ in the interval $[\xi_n,\xi_m]$. 
With this we arrive at
\be
 \dot{\rho} \: = \: 
 \sum_{n,m} \big( \sigma(\xi_m) - \sigma(\xi_n) \big) \, w_{nm}Q_m  \:
  - \: \frac{1}{2} \sum_{n,m}\limits \sigma^{\prime\prime}(\xi_{mn}) \, (\xi_m-\xi_n)^2 \, w_{nm} Q_m.
\ee{RHODOTTAYLOR}
The sum
\be
 \sum_{n,m} \big( \sigma(\xi_m) - \sigma(\xi_n) \big) \, w_{nm}Q_m,
\ee{VANISHSUM}
vanishes due to exchanging the summation indices only in the first term:
\be
 \sum_{n,m} \sigma(\xi_n) \left( w_{mn}Q_n - w_{nm} Q_m \right) \: = \: 0,
\ee{TRIVIALVANISH}
because of the total balance condition.
Since  $\sigma^{\prime\prime}>0$ the sign of $\dot{\rho}$ is hereby non-positive:
\be
 \dot{\rho} \: = \: - \frac{1}{2} \sum_{n,m}\limits \sigma^{\prime\prime}(\xi_{mn}) \, (\xi_m-\xi_n)^2 \, w_{nm} Q_m \: \le \: 0.
\ee{SIGNOFDOTRHO}
This result teaches us that by satisfying conditions 1 and 2 by $\sigma^{\prime\prime}>0$ and $\sigma(1)=0$
conditions 3 and 4 are automatically satisfied. The above sum (\ref{SIGNOFDOTRHO}) vanishes only if all
$\xi_m=\xi_n$. Since both $P_n$ and $Q_n$ are normalized, this is only possible if $P_n=Q_n$.

Finally we present the entropic divergence formula applied for the unidirectional grow and reset dynamics
using the classical ansatz $\sigma(\xi)=-\ln\xi$:
\be
 \dot{\rho} \: = \: 
  -\frac{1}{2} \sum_m\limits \frac{(\xi_m - \xi_{m+1})^2}{\xi_{m,m+1}^2} \, \mu_m Q_m \, 
  -\frac{1}{2} \sum_m\limits \frac{(\xi_m - \xi_{0})^2}{\xi_{m,0}^2} \, \gamma_m Q_m.
\ee{RHODOTFORGRMODEL}

\subsection{A treatable generalization}

In the previous subsection a proof by construction of a proper 
entropic divergence formula was given for general linear master equations
without assuming the detailed balance condition for the transition rates $w_{nm}$.
This result can easily be extended to dynamical
models based on nonlinear master equations of the type
\be
 \dot{P}_n \: = \: \sum_m\limits \left[ w_{nm}\, a(P_m) - w_{mn}\, a(P_n) \right]
\ee{NLINMASTER}
with $a(P)$ being a positive valued function. Practically one repeats the derivation
discussed above, with the important difference that now $\xi=a(P)/a(Q)$ is
the reference variable to be used.

%%%%%%%%%%%%%% copied from BiroSchramJenkovszky

The stationary distribution, $Q_n$, can be obtained by solving the total balance equation
\be
 0 \: = \: \sum_m\limits \left[w_{nm}\, a(Q_m) - w_{mn}\, a(Q_n) \right].
\ee{STACA}
We seek for an entropic divergence in the trace form
\be
 \rho[P,Q] \: \equiv \: \sum_n\limits  \sigma(P_n,Q_n) \: \ge \: 0,
\ee{ENTDISTA}
respecting conditions 1 and 2.
% with the triviality property $\rho[Q,Q]=0$ and otherwise positive. 
% The notation $\sigma_n(P_n)$ reminds that the value of this expression also depends on $Q_n$, not only on $P_n$.

The time derivative of this entropic distance is 
\be
\dot{\rho}=\sum_n\limits \pd{\sigma}{P_n} \cdot \dot{P}_n.
\ee{DOTRHOQP}
Using eq.(\ref{NLINMASTER}) it reads as
\be
 \dot{\rho} \: = \: \sum_{n,m}\limits \pd{\sigma}{P_n} \, \left[w_{nm}\, a(Q_m)\, \xi_m - w_{mn}\, a(Q_n)\, \xi_n \right].
\ee{RHODOTA}
In the second term of the above double sum we can perform the summation
over $m$ using the total balance (\ref{STACA}) %and with arbitrary $\lambda_n$, $\lambda_m$ arrive at
leading to
\be
 \dot{\rho} \: = \: \sum_{n,m}\limits
 \left[\pd{\sigma}{P_n} (\xi_m-\xi_n) \right] \, w_{nm} \, a(Q_m).
\ee{RHODOT2A}
We also can add a term,
\be
 \sum_{m,n}\limits 
  (\lambda_n-\lambda_m)  \, w_{nm} \, a(Q_m),
\ee{ADDINGZEROTERM}
which is zero. 
We utilize the total balance eq.(\ref{STACA}) in order to show that this term is indeed zero.
The proof is based on exchanging the summation indices $n$ and $m$ in the subtracted second term:
\be
 \sum_{n,m}\limits (\lambda_n-\lambda_m) \, w_{nm}\, a(Q_m) \: = \:
  \sum_n\limits \lambda_n \sum_m\limits \left[ w_{nm}\, a(Q_m) - w_{mn}\, a(Q_n) \right]  \: = \: 0.
\ee{ITISZERO}
Collecting all these results we rewrite the time derivative of the entropic divergence as
\be
 \dot{\rho} \: = \: \sum_{n,m}\limits
 \left[\pd{\sigma}{P_n} (\xi_m-\xi_n) \, + \, (\lambda_n-\lambda_m) \right] \, w_{nm} \, a(Q_m).
\ee{RHODOTATOTAL}
For satisfying the second law of thermodynamics (conditions 3 and 4) one aims to attend a definite sign to $\dot{\rho}$.
It is achieved by a similar construction as in the linear case. We set
\be
 \pd{\sigma}{P_n} \: = \: \mathfrak{s}^{\prime} (\xi_n),
\ee{IMPORTANT}
and choose $\lambda_n = \mathfrak{s}(\xi_n)$ in eq.(\ref{RHODOT2A}). 
With this the factor in the square brackets becomes
\be
 \mathfrak{s}^{\prime}(\xi_n) \, (\xi_m-\xi_n) + \mathfrak{s}(\xi_n) - \mathfrak{s}(\xi_m) \: = \:
 - \frac{1}{2} \mathfrak{s}^{\prime\prime}(\xi_{mn}) \, \left( \xi_m - \xi_n \right)^2,
\ee{SQBFAC}
and we can use again the remainder theorem for Taylor series in the Lagrange form. Again $\xi_{mn}$
is a value between $\xi_m$ and $\xi_n$, endpoints included. Our final result for the time derivative
of the entropic divergence is then summarized as follows:
\be
 \dot{\rho} \: = \: - \frac{1}{2} \sum_{n,m}\limits \mathfrak{s}^{\prime\prime}(\xi_{mn}) \,
 \left( \xi_m - \xi_n \right)^2 \, w_{mn} \, a(Q_m).
\ee{RHODOTFIN}
We conclude that the only requirement for $\dot{\rho} \le 0$ is that the function
$\mathfrak{s}(\xi)$ is subject to $\mathfrak{s}^{\prime\prime}(\xi) > 0$ for all arguments.
Having the  function $\mathfrak{s}(\xi)$ one reconstructs the entropic divergence 
as a sum of $\sigma(P_n,Q_n)$ terms by solving the partial differential equation
\be
 \pd{\sigma}{P_n} \: = \: \mathfrak{s}^{\prime} \left(\frac{a(P_n)}{a(Q_n)} \right).
\ee{SOLVETHIS}
Integration constants have to be set according to $\rho[Q,Q]=0$, i.e. $\sigma(Q_n,Q_n)=0$.
According to our best knowledge in this very general case we are the first to deliver such a proof.

\subsection{An example: $q$-generalization of the Kullback--Leibler divergence}

The classical result occurs for  linear dynamical models, $a(P)=P$, if one uses $\sigma(\xi)=-\ln \xi$.
The procedure described above arrives at the Kullback-Leibler divergence formula for the entropic distance.
\be
 \rho[P,Q] \: = \: \sum_n\limits Q_n \ln \frac{Q_n}{P_n}.
\ee{KLFORMULA}
Another example is given by the divergence using $a(P)=P^q$.  In this case we have
\be
\pd{\sigma}{P_n}=-(Q_n/P_n)^q, 
\ee{QPARC}
and the solution of eq.(\ref{SOLVETHIS}) with the proper integration constant leads to
\be
 \rho[P,Q] \: = \: \frac{1}{1-q} \left(1-\sum_n\limits Q_n^q P_n^{1-q} \right).
\ee{RENYIDIV}
This entropic divergence formula is of the special scaling trace formula 
\be
 \rho[P,Q] \: = \: \sum_n\limits Q_n f(P_n/Q_n),
\ee{RHOTYPE}
with the special function
\be
 f(x) \: = \: \frac{1-x^{1-q}}{1-q} \: = \: - \ln_q(x).
\ee{RATIOFUN}
Since $f^{\prime\prime}(x)=qx^{-q-1} > 0$ for $q > 0$ the Jensen inequality ensures properties 1 and 2.
In order to realize this we recall
\be
  \sum_n\limits Q_n f(P_n/Q_n) \: \ge \: f\left(\sum_n\limits Q_n \frac{P_n}{Q_n} \right) \: = \: f(1) = 0.
\ee{RHOJENSEN}
This means
\be
 \rho[P,Q] \: \ge \:  0.
\ee{RHONONNEG}
Recalling the formula for the Tsallis entropy,
\be
 S_T[Q] \: = \: \frac{1}{q-1} \sum_n\limits \left( Q_n - Q_n^q \right),
\ee{TSALENTROP}
one obtains $S_T[U]= (W^{1-q}-1)/(1-q)$ for the uniform distribution. Now the entropic
divergence between the stationary distribution, $Q$ and
the uniform distribution, $U_n=1/W$ for $n=1,2,\ldots W$ can be expressed as
\be
 \rho[U,Q] \: = \: W^{q-1} \, \Big\{ \, S_T[U] - S_T[Q]  \,  \Big\}.
\ee{URENYIDIV}
Here $\rho[U,Q] \ge 0$ ensures that among all possible stationary distributions, $Q_n$, the uniform
distribution has the maximal Tsallis entropy, i.e. $S_T[U] \ge S_T[Q]$.

This properly constructed entropic divergence is proportional
to the difference of Tsallis entropies (not the R\'enyi ones) for comparing the uniform distribution
with the stationary one. The proportionality constant, $W^{q-1}$, in eq.(\ref{URENYIDIV}) 
can be melted with the definition of $\rho[U,Q]$.
This is  an argument in favor of Tsallis entropy instead of the R\'enyi formula.

In this subsection we have shown two examples of analytic expressions for the entropic divergence.

%%%%%%%%%%%%%% end of copy

\subsection{When detailed balance is necessary}

%Following the above survey of possible entropic distance formulae and their behavior relative to corresponding
%individual entropy expressions, the question arises whether more complex dynamics as discussed here
%also may allow for, or even lead to, the use of entropic divergences beyond the Kullback--Leibler form.
%This might also establish the use of so called deformed entropies deviating from the the additive behavior
%inherent in the Boltzmann and in the R\'enyi constructions. After all Boltzmann's H-theorem is given for
%pairwise interactions, nevertheless using the detailed balance condition for the transition rates between
%momentum states of the colliding pairs.

Here we study the case when the transition probability from one state of the system
to another depends on both the initial and final state occupation probabilities. 
Our above presented proof now fails, and only
detailed balance conditioned elementary rates allow for a definite sign of the change of entropic distances.
Already for a factorized dependence the detailed balance is necessary.

We consider here the dynamical equation
\be
 \dot{P}_n \: = \: \sum_m\limits \left[w_{nm}a(P_m)\, b(P_n)-w_{mn}a(P_n)\, b(P_m) \right] 
 %          \: \equiv \: \sum_m\limits \left( A_{nm} - A_{mn} \right),
\ee{ABDYNAMIC}
together with trace form quantity
\be
 \rho[P,Q] \: \equiv \: \sum_n\limits \sigma(P_n,Q_n).
\ee{ABDEFH}
The change of this quantity in time is given as
\be
 \dot{\rho} \: = \: \sum_n\limits \pd{\sigma}{P_n} \, \dot{P}_n.
\ee{ABCHANGEH}
Now we use the index antisymmetry inherent in eq.(\ref{ABDYNAMIC}) to obtain
\be
 \dot{\rho} \: = \: \frac{1}{2} \sum_{n,m}\limits w_{nm} a(P_m) b(P_n)
 \left[\pd{\sigma}{P_n}-\pd{\sigma}{P_m} \right] \cdot 
 \left[1-\frac{w_{mn}a(P_n)\, b(P_m)}{w_{nm}a(P_m) \, b(P_n)} \right].
\ee{ABHDOT}
Due to the mixed dependence on $n$- and $m$-indexed quantities the monotonicity or second derivative
argumentation, presented in the previous subsections, cannot be carried out now.
However, assuming that the transition rates, $w_{nm}$ and $w_{mn}$ satisfy the 
{\em detailed balance condition} with the stationary distribution, 
\be
 \frac{w_{mn}}{w_{nm}} \: = \: \frac{b(Q_n) \, a(Q_m)}{a(Q_n) \, b(Q_m)},
\ee{ABDETBAL}
we obtain
\be
 \dot{\rho} \: = \: \frac{1}{2} \sum_{n,m}\limits w_{nm} a(P_m) \, b(P_n) \,
 \left[\pd{\sigma}{P_n} - \pd{\sigma}{P_m} \right] \cdot
 \left[ 1 - \frac{f(P_n)/f(Q_n)}{f(P_m)/f(Q_m)} \right]
\ee{ABHDOTDETBAL}
with the general function $f(x)=a(x)/b(x)$. The above result can be rewritten in terms of the original
rates as
\be
 \dot{\rho} \: = \: \frac{1}{2} \sum_{n,m}\limits w_{nm} b(P_n) \, b(P_m) \, f(Q_m) \,
 \left[\pd{\sigma}{P_n} - \pd{\sigma}{P_m} \right] \cdot
 \left[ \frac{f(P_m)}{f(Q_m)} - \frac{f(P_n)}{f(Q_n)} \right].
\ee{ABDETBALDOT}
From here it is obvious that one shall use the following ratio variable
\be
 \xi_n \: = \: \frac{f(P_n)}{f(Q_n)} \: = \: \frac{a(P_n)}{a(Q_n)} \, \frac{b(Q_n)}{b(P_n)}.
\ee{ABDETBALXI}
For a definite sign, $\dot{\rho}\le 0$ one needs that
\be
 \pd{\sigma}{P_n} \: = \: \mathfrak{s}^{\prime}(\xi_n)
\ee{ABSIGMADIFFEQ}
to obtain
\be
 \left[\mathfrak{s}^{\prime}(\xi_n) - \mathfrak{s}^{\prime}(\xi_m) \right] \cdot \left[ \xi_m - \xi_n \right] \: \le \: 0.
\ee{ABDETBALSPRIMEXI}
This expression, based again on the remainder theorem for Taylor series can be expressed using
$\mathfrak{s}^{\prime\prime}(\xi_{nm})$ at an intermediate point. Finally we arrive at
\be
 \dot{\rho} \: = \: - \frac{1}{2} \sum_{n,m}\limits w_{nm} b(P_n) \, b(P_m) \, f(Q_m) \,
 \mathfrak{s}^{\prime\prime}(\xi_{nm}) \, (\xi_m-\xi_n)^2.
\ee{ABRHODOTFINAL}
Global stability is achieved for any $\mathfrak{s}^{\prime\prime}>0$ function. 
This is the same result as in the previous subsection eq.(\ref{RHODOTFIN}), but only 
subject to the detailed balance condition.

Again, if $a(P)\ne P$ and $b(P) \ne 1$, the classical choice, $\mathfrak{s}(\xi)=-\ln \xi$, does not
lead to the Kullback--Leibler form in terms of the probabilities $P_n$ and $Q_n$.
In such cases $\xi_n \ne P_n/Q_n$. However a modified connection to the entropy formula
also occurs in this general treatment. With the standard logarithmic $\mathfrak{s}(\xi)$ function
one has
\be
 \pd{\sigma}{P} \: = \: - \frac{1}{\xi} \: = \: - \frac{f(Q)}{f(P)}.
\ee{SIGDERLOGS}
Its proper solution utilizes the primitive function, $g(x)=\int \frac{dx}{f(x)} $:
\be
 \sigma(P,Q) \: = \: f(Q) \Big( g(Q) - g(P) \Big) \: = \: \frac{g(Q)-g(P)}{g^{\prime}(Q)}.
\ee{SIGLOGS}
Using the Taylor series remainder theorem again up to the second derivative
the entropic distance equals to the following expression
\be
 \rho[P,Q] \: = \: - \frac{1}{2} \, \sum_n\limits \frac{g^{\prime\prime}(R_n)}{g^{\prime}(Q_n)} \, (P_n-Q_n)^2
\ee{RHOLOGSGFORM}
with $R_n\in[Q_n,P_n]$ being a certain value between $P_n$ and $Q_n$. Finally using the function $f(x)$
instead of the derivatives of $g(x)$ we arrive at the expression
\be
 \rho[P,Q] \: = \: \frac{1}{2} \, \sum_n\limits \frac{f^{\prime}(R_n)}{f^2(R_n)} \, f(Q_n) \, (P_n-Q_n)^2.
\ee{RHOLOGSFFORM}
This expression is positive for all dynamics using $f^{\prime} > 0$ for unequal $P$ and $Q$ distributions
and zero only if they coincide. Common choices, like $a(P)=P$, $b(P)=1+\lambda P$, satisfy this condition.
Physical values for $\lambda$ are bigger than $-1$.

The distance between the uniform and the stationary distribution is given by
\be
 \rho[U,Q] \: = \: Z[Q] \, \Big(S[U]-S[Q] \Big)
\ee{RHOLOGSUQ}
with the entropy definition
\be
 S[Q] \: \equiv \: - \frac{1}{Z[Q]} \, \sum_n\limits f(Q_n) g(Q_n)
\ee{ENTROPLOGSDEF}
and the distribution-dependent sum
\be
 Z[Q] \: \equiv \: \sum_n\limits f(Q_n).
\ee{ZLOGSDEF}
As eq.(\ref{RHOLOGSUQ}) clearly shows the entropic distance in this more general case cannot be
interpreted as relative entropy due to the $Q$-dependence of the prefactor.

Finally closing this subsection we note that a non-factorizing dependence of the transition probability
on the starting and final occupation probabilities does not allow for such a proof even when detailed balance
is imposed.

%%%%%%%%%%

%%%%%%%%%%%%% STAT DIST %%%%%%%%%%%%%%

\section{Stationary Distributions}

%\textcolor{Green}{
%\begin{itemize}
%\item Transition rates $\rightarrow$ $Q_n=\lim_{t\to\infty}\limits P_n(t)$
%\item Stationary distribution $\rightarrow$ relation between rates
%\item Large $n$ limit: probability flow equation
%\item Some known distributions: Exponential, Pareto, Gompertz, gamma, Gauss, log-normal
%\end{itemize}
%}

We discuss separately the discrete and the continuous state evolutions and the corresponding
stationary distributions. Several well-known statistical distributions are successfully reconstructed
in the framework of this unified mathematical model.

\subsection{Discrete state space processes}

In order to identify the stationary distributions, $Q_n$, one considers $\dot{P}_n=0$  and solves
the evolution equation:
\be
 0 \: = \: \sum_m\limits \left[ w_{nm} Q_m - w_{mn} Q_n \right].
\ee{SQSTAT}
For unidirectional growth with resetting  for any $n > 0$  we obtain a simple recursion
\be
 \mu_{n-1}Q_{n-1} \: = \: (\mu_n+ \gamma_n) Q_n.
\ee{QRECURN}
The $n=0$ state has to be handled carefully.  $Q_0$ can be obtained either from
the normalization condition, $\sum_n\limits Q_n = 1$, or using eq.(\ref{PDOTUNIRESET}) for $n=0$
\be
 0 \: = \: \sum_{n=0}^{\infty}\limits \gamma_n \, Q_n \: - \: (\mu_0+\gamma_0) \, Q_0.
\ee{QZERO}
These alternative ways always lead to the same result. 
This nontrivial statement was proven for semi-infinite chains in \cite{BiroNedaPRE2017}.
The resolution of this recursion is finally given by
\be
 Q_n \: = \: \frac{\mu_0Q_0}{\mu_n} \, \prod_{j=1}^n\limits \left( 1 + \frac{\gamma_j}{\mu_j} \right)^{-1}.
\ee{QNFORUNIRESET}
Let us  discuss now particular transition rates. First we consider
constant (initial state-independent) growth and resetting rates, $\mu_n=\mu$, $\gamma_n=\gamma$.
The recursion equation (\ref{QNFORUNIRESET}) delivers the $n$-th power of the same term:
\be
 Q_n \: = \: Q_0 \left(1+\frac{\gamma}{\mu} \right)^{-n}.
\ee{QNCONCON}
This is the {\em geometrical distribution}, which is easily transformed to an exponential,
\be
 Q_n \: = \: \frac{\gamma}{\mu+\gamma} \, \eon{-n \cdot \ln(1+\gamma/\mu)}.
\ee{QNCONCONEXP}
In the case when $n$ denotes the number of energy quanta, $E_n=n\epsilon$, 
this formula reproduces the Boltzmann distribution with the temperature 
\be
 k_BT \: = \: \frac{\epsilon}{ \ln(1+\gamma/\mu) }.
\ee{BOLTEMP}
For non-thermic applications, one can view this quantity as a generalized temperature.
In several practical applications the reset rate is much smaller than the growth rate,
$\gamma \ll \mu$, simplifying the above notion of the temperature:
\be
 k_B T \: \to \: \epsilon \, \frac{\mu}{\gamma}.
\ee{GENTEMLIMIT}

More exciting is to consider linear preference rates for the local transition, $\mu_n=\sigma(n+b)$.
This reflects the Matthew principle: ''For whosoever hath, to him shall be given,\ldots''
\cite{Matthew}. 
In our mathematical model the growth rate linear in $n$ means that the next unit will be added
sooner to those who have already more. As a result of this, in a given time they gain more then the
others.
A state $n$ in this respect denotes having $n$ units of some arbitrary goods 
(energy, money, network connection, past citations) and $b>0$ is
a threshold parameter. Very often $b=1$ is chosen, setting $\mu_0=\sigma$.
The stationary distribution in this case is a ratio of $n$-fold products with different offsets,
\be
 Q_n \: = \: Q_0 \, \prod_{j=1}^n\limits \frac{j-1+b}{j+b+\gamma/\sigma}
   \: = \: Q_0 \, \frac{(b)_n}{(c)_n},
\ee{QNUNILIN}
with $c=b+1+\gamma/\sigma$.  Here we use the Pochhammer symbol, $(b)_n=b(b+1)\ldots(b+n-1)$,
a generalization of the factorial. (Indeed $(1)_n=n!$.) From the normalization,
\be
 \sum_n\limits Q_n \: = \: Q_0 \, \sumi{n} \frac{(b)_n}{(c)_n} \: = \: 1,
\ee{NORMALIZEWARING}
follows the final result for the stationary distribution of processes with linear preference rates:
\be
 Q_n \: = \: \frac{c-1-b}{c-1} \, \frac{(b)_n}{(c)_n}.
\ee{QNWARING}
This is the {\em Waring distribution}, and has been considered in failure processes
by Irwin \cite{Irwin,Irwin2}.
The high-$n$ tail of this distribution is a power-law:
\be
 Q_n  \xrightarrow{n\to\infty} \, \frac{\gamma}{\gamma+\sigma b} \, \frac{\Gamma(c)}{\Gamma(b)} \: n^{-1-\gamma/\sigma}.
\ee{QWARINGHIGHN}
In particular with closely vanishing resetting rates, $\gamma\to 0$ one reconstructs from this result
the {\em Zipf distribution} \cite{ZIPF},
\be
 Q_n \: \sim \:  n^{-1}.
\ee{ZIPFDIST}

\subsection{Continuous state space}
\label{CONTDIST}

Concentrating to the large $n$ behavior of such stationary distributions, a continuous variable limit,
$n \to \infty$ while $x=n \Delta x$ kept finite, is of special interest. In this limit both the evolution
equation and the determination of the stationary distribution are simpler. 
In order to gain a nontrivial equation for the unidirectional step-up evolution with long jumps
to the zero state, one realizes that the short jump and long jump coefficients have to scale differently.
We define the continuous rate functions, $\mu(x)$ and $\gamma(x)$, respectively as to satisfy
\be
 \mu(x) \: \equiv \: \mu(n \cdot \Delta x) \: = \: \mu_n \, \Delta x, \qquad \mathrm{and} \qquad
 \gamma(x) \: \equiv \: \gamma (n \cdot \Delta x) \: = \: \gamma_n. 
\ee{CONTRATES}
We arrive at a flow-like continuous state space master equation
\be
 \pd{}{t} {\cal P}(x,t) \: = \: - \pd{}{x} \Big( \mu(x) \, {\cal P}(x,t) \Big) \, - \, \gamma(x) \, {\cal P}(x,t).
\ee{MASTERFLOW}
This equation essentially differs from the Fokker--Planck equation (\ref{CONTIDIFFUSEQ}): the reset process
leads to a genuinely new term.

\newcommand{\infi}{\int_0^{\infty}\limits }
The stationary probability density function (PDF) in this case is given by
\be
 {\cal Q} (x) \: = \: \frac{\mu(0)\, {\cal Q}(0)}{\mu(x)} \: \eon{-\int_0^x\limits \frac{\gamma(u)}{\mu(u)} \, du}.
\ee{STATIFLOW}
Here ${\cal Q}(0)$ is obtained from the condition $\infi {\cal Q}(x) dx \: = \: 1$.

We consider again simple rates, starting with $\gamma(x)=\gamma$, a state independent resetting rate.
For constant growth rate, $\mu(x)=\mu$, we regain the {\em exponential distribution} (\ref{QNCONCONEXP}) in the 
$\gamma_n \ll \mu_n$ limit:
\be
 {\cal Q}(x) \: = \: \frac{\gamma}{\mu} \: \eon{ -\frac{\gamma}{\mu} \, x}.
\ee{CONTEXP}
For a linear preference in the growth rate, $\mu(x)=\sigma(x+b)$, the stationary PDF is the 
{\em Tsallis--Pareto distribution},
\be
 {\cal Q}(x) \: = \: \frac{\gamma}{\sigma b}  \, \left(1 + \frac{x}{b} \right)^{-1-\gamma/\sigma}. 
\ee{CONTSALPARETO}
In this view the Tsallis--Pareto distribution is the continuous limit of the Waring distribution.
The growth preference can also be a nonlinear function. 
Modelling human mortality an exponentially increasing rate
has been considered by Benjamin Gompertz \cite{CANCER,BGG}. For 
$\mu(x)=\sigma \eon{bx}$ eq.(\ref{STATIFLOW}) delivers the {\em Gompertz distribution}:
\be
 {\cal Q}(x) \: = \: \frac{\gamma/\sigma}{1-\eon{- \gamma/b\sigma}} \, \eon{-bx - \frac{\gamma}{b\sigma}\left(1-\eon{-bx}\right)}.
\ee{CONTGOMPERTZ}
Indeed the structure of the stationary PDF, suggested by eq.(\ref{STATIFLOW}), for constant
resetting rate, $\gamma(x)=\gamma$, fits in the general scheme used in modeling 
failure processes related to technical and health insurance studies \cite{FAILUREHAZARD,Xekalaki}. 
For constant $\gamma$ the structure of the stationary PDF is always of the form
\be
 {\cal Q}(x) \: = \: \frac{ h(x) \, \eon{-H(x)}}{\eon{-H(0)} - \eon{-H(\infty)} },
\ee{HAZARDPDF}
with $h(x)=H^{\prime}(x)$ being the hazard rate and $H(x)$ itself the cumulative hazard. 
%For proper normalization $H(0)=0$ and $\lim_{x\to\infty}\limits H(x) = \infty$ has to be fulfilled.
The survival rate can be expressed by the decreasing exponential of the cumulative hazard:
\be
 R(x) \: = \: \int_x^{\infty}\limits\!{\cal Q}(u) \, du \: = \: \frac{\eon{-H(x)}-\eon{-H(\infty)} }{\eon{-H(0)} - \eon{-H(\infty)}}.
\ee{SURVIVAL}
In the special case $H(0)=0$ and $H(\infty)=\infty$ one arrives at $R(x)=\exp(-H(x))$.
The growth rate and the constant resetting rate are related by the hazard rate
\be
 \mu(x) \, h(x) \: = \: \gamma.
\ee{RESETHAZARD}

%%%%%%%%%%%%%%%%%%% WE ARE HERE

\subsection{Generalized fluctuation--dissipation relation}

Abandoning the constancy of the resetting rate, a state dependent $\gamma(x)$ and $\gamma_n$
scenario should be envisioned. Is there some compact relation between $\mu$ and $\gamma$ also in this
case? The answer is affirmative:  from the formula (\ref{STATIFLOW}) for the stationary PDF it
is straightforward to derive that the growth rate is in general equal to a truncated
expectation value of the resetting rate at stationarity:
\be
 \mu(x) \: = \: \frac{1}{{\cal Q}(x)} \, \int_x^{\infty}\limits\!\gamma(u)\, {\cal Q}(u) \, du.
\ee{MUGAMQ}
This result resembles the generalization of the fluctuation--dissipation relation, obtained
by investigating a colored noise Langevin equation 
\cite{BiroGyorgyietal2004,BiroJakovacPRL2005,BiroJakoGyorgyiQM2005,BiroJakoGyorgyiSQM2005}.

%%%%%%%%%%%%%%%%%%%%%
In order to support the above statement we briefly present here the main idea.
In that model the Langevin equation,
\be
 \dot{p} + (g p - \xi) \: = \: 0,
\ee{LANGEVIN}
for the momentum $p$ of a particle moving on a line is supported by two stochastic coefficients: $\xi$
represents the noisy pushing force while the $g$ damping coefficient also contains a stochastic part.
Altogether the average total force factor depends on the particles' momentum, $p$:
$\exv{g p - \xi} = \gamma(p)$ and the Markovian (short-time) self-correlation of the noisy force
contains another $p$-dependence
\be
 \exv{(g p - \xi)(t) \, (g p - \xi)(t^{\prime})   } \: = \: \delta(t-t^{\prime}) \, \mu(p).
\ee{NOISYFORCECORREL}
The corresponding Fokker--Planck equation then includes these $p$-dependent (colored) noise functions:
\be
 \pd{}{t} {\cal P} \: = \: \pd{}{p} \big(\gamma(p) {\cal P} \big)  \, + \, \pd{^2}{p^2} \big(\mu(p) {\cal P} \big). 
\ee{FOKKERPLANCK}
The resulting stationary distribution is in a similar form than eq.(\ref{STATIFLOW}),
\be
 {\cal Q}(p) \: = \: \frac{{\rm const.}}{\mu(p)} \: \eon{-\int_0^p\limits\!\frac{\gamma(u)}{\mu(u)} \, du}.
\ee{STATFPLANCK}
The fluctuation--dissipation relation becomes (\ref{MUGAMQ}).

%%%%%%%%%%%%%%%%%%%%

In the followings we turn our attention to some specific cases.
For the exponential distribution ${\cal Q}(x) \sim \exp(-x/T)$  combined with a constant
$\gamma(x)=\gamma$ one arrives at
\be
 \mu(x) \: = \: T \, \gamma,
\ee{CONTEXPMUGAMQ}
resembling the classical Einstein--Kubo formula.

Finally we present the discrete version of this generalized fluctuation--dissipation relation. This is easy to
derive by summing up eq.(\ref{QRECURN}) from the index $n+1$ to infinity,
\be
 \mu_n \: = \: \frac{1}{Q_n} \, \sum_{m=n+1}^{\infty}\limits \gamma_m Q_m.
\ee{DISCRETEFDT}
For the exponential distribution, $Q_n = \eon{-\beta \, n \epsilon}/Z$, and constant resetting rate, $\gamma_n=\gamma$,
one obtains the quantum Kubo formula
\be
 \mu_n \: = \: \frac{\gamma}{\eon{\beta \, \epsilon}-1}.
\ee{QKUBO}
Before closing this subsection we summarize the resetting and growth rates 
leading to the most commonly observed stationary PDF-s (Table\ref{TAB1}).

\renewcommand{\arraystretch}{1.5}

\begin{table}[h]
\label{OURTABLE}
\begin{center}
\begin{tabular}{||c|c|c||}
\hline \hline
 $\gamma(x)$ &  $\mu(x)$  &  $Q(x)$  \\ \hline \hline
 $\gamma$ & $\mu$  &  Exponential: $\sim \eon{-(\gamma/\mu) x}$ \\ \hline
 $\gamma$ & $\sigma(x+b)$  & Tsallis--Pareto: $\sim (1+x/b)^{-1-\gamma/\sigma} $\\ \hline
 $\gamma$ & $\sigma x^{\alpha}$, $\alpha<1$  & Weibull: $\sim x^{-\alpha} \eon{-bx^{1-\alpha}}$  \\ \hline
 $\gamma$ & $\sigma(x+a)(x+b)$ & Pearson: $\sim (x+a)^{-1-v}(x+b)^{-1+v}$ \\ \hline
 $\gamma$ & $\sigma \eon{x}$  & Gompertz: $\sim \exp{\Big(\frac{\gamma}{\sigma}\eon{-x} - x\Big)}   $ \\ \hline 
 $\ln (x/a)$ & $\sigma x$ & Log-Normal: ${\cal Q}(x) \, dx \: \sim \: \eon{-\gamma^2/2\sigma} \, d\gamma $ \\ \hline
 $ x$ & $\sigma^2$ & Gauss: $\sim \eon{-x^2/2\sigma^2}   $ \\ \hline
 $\sigma(ax-c)$ & $\sigma x$ & Gamma: $\sim x^{c-1} \, \eon{-ax}  $ \\ \hline
\hline
\end{tabular}
\end{center}
\caption{\label{TAB1} Resetting and growth rates for the most common stationary PDF-s.}

\end{table}

\renewcommand{\arraystretch}{1.0}

%It is worth noting that for $\gamma(x)=\gamma$ constant only the ratio, $\mu(x)/\gamma$,
%governs the stationary distribution.

\subsection{Evolution towards the stationary value}

The time evolution starting from a generic initial distribution, $P_n(0)$,  can be 
very complicated. It is more transparent to study the evolution of the ratio
of continuous PDF-s to their stationary counterparts. We define 
\be
 \xi(x,t) \: = \: \frac{{\cal P}(x,t)}{{\cal Q}(x)}.
\ee{XIDEF}
Substituting ${\cal P} = \xi {\cal Q}$ into  eq.(\ref{MASTERFLOW}) one obtains
\be
 {\cal Q} \pd{\xi}{t} \: = \: -\xi \pd{}{x} \big( \mu \, {\cal Q} \big) - \mu \, {\cal Q} \pd{\xi}{x} \, - \; \gamma \xi \, {\cal Q}.
\ee{XIQMASTER}
Using the stationarity condition (\ref{STATIFLOW}) the first and third term on the right hand side of this equation cancel
each other.
The remaining equation can be divided by ${\cal Q}(x)$ for all $x$ where $Q(x)\ne 0$. This results in
\be
 \pd{\xi}{t} + \mu(x) \pd{\xi}{x} \: = \: 0.
\ee{XIMASTER}
The above equation describes a flow  with the  velocity field $\mu(x)$. 
The evolution of the local
ratio of the actual to the stationary PDF is independent of the resetting rate, $\gamma(x)$. 
The resetting rate was important in
shaping the stationary PDF, ${\cal Q}(x)$, but it plays no direct role in the evolution of this ratio.
Furthermore, the time evolution of $\xi(x,t)$ is soliton-like. We consider the $\xi$-flow in the form
\be
 \pd{\xi}{t} \: = \: - \mu(x) \pd{\xi}{x} \: = \: - \pd{\xi}{y},
\ee{XISOLI}
with 
\be
 y(x) \: \equiv \: \int_0^x\limits\! \frac{du}{\mu(u)}.
\ee{YDEF}
The solution for $\xi$ is a function that depends only on  $y-t$, describing a general 
solitary wave  propagating  along $x$ as driven by %the step-up transition rate,
$\mu(x)$:
\be
 \xi(x,t) \: = \: f(y(x)-t) \: = \: \xi(x_t,0)
\ee{XISOLUTION}
with $dx_t/dt=-\mu(x_t)$ defining characteristic trajectories. For constant $\mu(x)=\mu$ one has
$\xi(x,t)= \xi(x-\mu t,0)$ a shift with constant velocity. With the natural boundary condition
setting $\xi(0^{-},t)=1$, the deviation from the stationary PDF moves with
constant velocity towards higher $x$. For growth rates with linear preference, $\mu(x)=\sigma(x+b)$,
we obtain:
\be
 \xi(x,t) \: = \: \xi( x\eon{-\sigma t} - b(1-\eon{-\sigma t}), 0).
\ee{LINPREFXI}
%For sufficiently late time ending up with $\xi$-ratios at negative $x$ for $b>0$, which are set to one, $\xi(-b,0)=1$.
In this case the deviation wave  travels with an accelerating pace.

The lesson learned from this subsection is that the approach to the stationary PDF is independent
of the resetting rate. We have given a compact equation with soliton-like solution for
this evolution.

Instead of discussing more examples we turn in the next section to the problem of stability for the stationary PDF-s.
We are particularly interested in a control quantity, called entropic divergence, between the actual $P_n(t)$ and the
stationary $Q_n$, that can only shrink during the evolution.

\subsection{Entropic divergence for the growth and resetting process}

Finally, before turning to applications in complex systems, 
let us briefly discuss some specialties of the entropic divergence measure
for the unidirectional growth and reset process.

Now we use a symmetrized entropic distance measure, $\kappa(\xi)$, for the growth and reset
dynamics.
Since the ratio $\xi(x,t)={\cal P}(x,t)/{\cal Q}(x)$, is always positive, $\kappa(\xi)$
is non-negative.
The continuous entropic divergence is defined by the following integral
\be
 \rho[{\cal P}, {\cal Q} ] \: = \: \inti \kappa(\xi(x,t)) \, {\cal Q}(x) \, dx \ge 0.
\ee{CONTENTDIST}
It evolves according to the evolution of $\xi(x,t)$. From eq.(\ref{XIMASTER}) it follows that
\be
 \pd{\kappa}{t} \: = \: - \mu(x) \, \pd{\kappa}{x}.
\ee{SEVOL}
Using this equality the time derivative of the entropic distance is given by
\be
 \dot{\rho} \: = \: \inti \pd{\kappa}{t} {\cal Q}(x) \, dx
 \: = \: - \inti \mu(x) \, {\cal Q}(x) \, \pd{\kappa}{x} \, dx.
\ee{RHODOTSTEPUP}
Partial integration by $x$, taking $\xi(0,t)=1$ and therefore $\kappa(\xi(0,t))=0$  leads to
\be
 \dot{\rho} \: = \:  \inti \kappa(\xi(x,t)) \,  \pd{}{x} \left( \mu(x) {\cal Q}(x)\right) \, dx.
\ee{DOTRHOQDERIV}
Finally using the stationary solution eq.(\ref{STATIFLOW}), we arrive at an expression
which ensures the shrinking of the entropic distance to the stationary PDF for $\gamma(x) > 0$:
\be
 \dot{\rho} \: = \: - \inti \kappa(\xi(x,t)) \, \gamma(x) \, {\cal Q}(x) \, dx \: \le \: 0.
\ee{RHODOTSHRINKS}
What is left to be proven is that this distance actually shrinks as long as the stationary PDF is achieved.
For this purpose we use again the Jensen inequality,
\be
 \int\!p(x) \kappa(\xi(x,t)) \, dx \: \ge \: \kappa \, \Bigg(\int\!p(x)\xi(x)\, dx \Bigg),
\ee{CONTJENSEN}
for $\kappa^{\prime\prime}>0$,
with an arbitrary PDF, $p(x)\in [0,1]$, normalized as $\inti p(x)\, dx = 1$. We choose for our
purpose the following ''escort type'' probability density
\be
 p(x,t) \: = \: \frac{\gamma(x) \, {\cal Q}(x)}{\inti \gamma(u) \, {\cal Q}(u) \, du}.
\ee{OURp}
This construction results in the following bound on the shrinking rate of the symmetrized entropic
distance
\be
 \dot{\rho} \: \le \: - \exv{\gamma}_{\infty} \, \kappa\left( \frac{\exv{\gamma}_t}{\exv{\gamma}_{\infty}} \right).
\ee{SHRRATE}
Here we utilized the notation
\be
 \exv{\gamma}_t \: \equiv \: \inti \gamma(u) \, {\cal P}(u,t) \, du,
\ee{GAMEXPATT}
and
\be
 \exv{\gamma}_{\infty} \: \equiv \: \inti \gamma(u) \, {\cal Q}(u) \, du,
\ee{GAMEXQ}
From this result it is straightforward to see that $\dot{\rho} < 0$ until the stationary PDF is achieved.
At this instant $\dot{\rho}=0$ due to $\kappa(1)=0$.

%%%%%%%%%%%%%%% APPLICATIONS %%%%%%%%%%

\section{Applications}

%\textcolor{Brown}{
%\begin{itemize}
%\item Network: Degree Distributions
%\item Citation Fraction with Exponential Background Growth
%\item Hadron Multiplicity
%\item Income and Wealth
%\item Sample Space Reducing Processes
%\item Biodiversity Reset Time Distribution
%\end{itemize}
%}

In this section we review a few applications of the unidirectional sustained random growth
process, described in eqs.(\ref{PDOTUNIRESET}) and (\ref{MASTERFLOW}). 
Although the data we analyze usually represent discrete distributions,
for the sake of simplicity we discuss them in the continuous model's  framework.
The stationary distributions in the continuous limit, ${\cal Q}(x)$, are of simpler form.

This type of stochastic dynamics, together with its stationary distribution, is not uncommon:
very often a quantity $x$ changes by a small amount in one step, and only in the growth direction. 
The low probability resetting event to the reference state $x=0$ makes
such processes nontrivial, leading to a rich variety of stationary distributions at the end.

% In many systems the resetting is realized by an exponential dilation of the probability.
In many systems the resetting is realized by an exponential dilatation of the sample space.
Let the non-normalized density for the observed quantity, $w(x,t)$, grow in time as
\be
 \pd{w}{t} \: = \: - \pd{}{x}(\mu w).
\ee{WPDFGROW}
This is obviously a one-directional flow. We are interested in the normalized probability density,
\be
 {\cal P}(x,t) \: \equiv \: \frac{1}{Z} \, w(x,t)
\ee{PROBDENSW}
with
\be
 Z \: = \: \inti w(x,t) \, dx.
\ee{ZSUMFORW}
The corresponding evolution equation for the PDF writes as
\be
 \pd{}{t} {\cal P}(x,t) \: = \: - \pd{}{x} \Big( \mu(x) \, {\cal P}(x,t) \Big) \, - \, \frac{\dot{Z}}{Z} \, {\cal P}(x,t).
\ee{PDFEQW}
Comparing with eq.(\ref{MASTERFLOW}) we obtain the gamma factor
\be
 \gamma \: = \: \frac{\dot{Z}}{Z} \, 
\ee{GAMMADEFW}
which for exponentially growing systems is a constant. Therefore the results derived in section
\ref{CONTDIST} can be directly applied.

This type of sustained random growth
%, where the inflow is equivalent with the sum of all resetting
%transitions due to the conservation of the norm, 
can be made equivalent to the sampling space reduction
process family, cf. \cite{ThurnerSSR}, based on a correspondence between transition rates
leading to the same stationary distribution  \cite{Thurner2017arxiv}.

Our list of examples identifies the physical meaning of state labels $n$ in the data and $x$
in the continuous model, the elementary transition rates, $\gamma$ and $\mu(x)$,
and then compares the stationary distribution classified above with findings in simulations and measurements
from the literature. 
By our restricted set of examples we are not aiming at completeness. 
By the richness of literature and examples of power-law tailed and other
semi-exotic statistics it would be an impossible mission.

%%%%%%%%%%%% HERE OCT 5

\subsection{Degree distribution in networks}

Our first example is already a classical one: the degree distribution, the number of connections from and to a
node, in many large networks shows a stretched exponential tailed statistics.
It includes the pure exponential and the  power-law tails \cite{EMERG} as particular cases. 
Seemingly this is true both for directed and undirected networks \cite{BarabasiREVMOD}.
In the following we speculate on how such distributions could follow from our general master equation
presented above.

\begin{figure}[hbt]
\begin{center}
\includegraphics[width=0.9\textwidth]{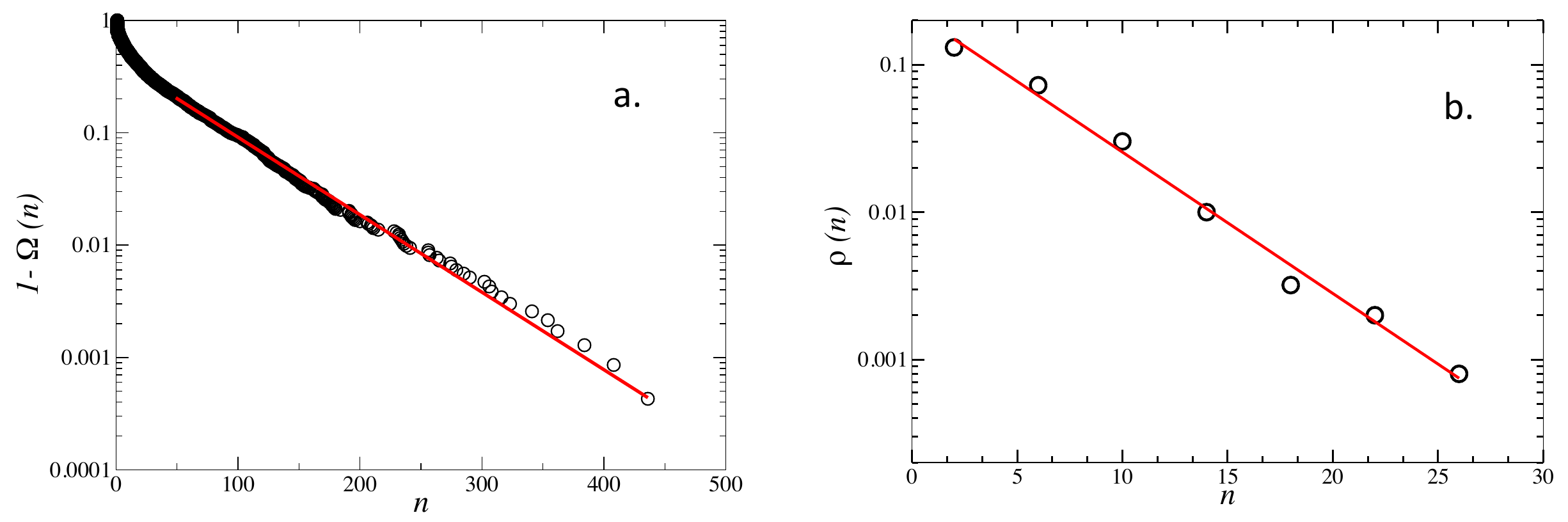}
\end{center}
\caption{\label{EXPONET}
 Exponential degree distributions in educational networks. The left figure shows results for the cumulative 
 distribution function $1-\Omega(k)$ for the
 ERASMUS student exchange networks. The right panel shows results for the logarithmically binned 
 $\rho(k)$ in-degree distribution function for the collaboration network of the Hungarian Talent Supporting Organizations.}
\end{figure}

The simplest growth model for a network is characterized by a constant growth rate $\mu$ and a constant resetting
rate $\gamma$. This leads to an exponential degree distribution at stationarity. 
Such distributions have been observed in world-wide marine transportation networks,
in e-mail networks, power grid networks and educational collaboration networks \cite{WeibingDeng}.
For these complex networks the evolution dynamics does not seem to show any preference. 
We exemplify this simple case first by the degree distribution for the ERASMUS collaboration network of European universities studied by us earlier \cite{ERASMUS}, which shows a clear exponential tail.  In order to illustrate this in the left panel 
of Fig.\ref{EXPONET} we plot $1-\Omega(n)$, with $\Omega(n)$ the cummulative distribution function: 
$\Omega(n)=\int_0^n \rho(x)dx$. The exponential nature of this, suggests that $\rho(n)$ is also exponential. 
The degree distribution of this network was constructed from an exhaustive dataset available for the year 2003, 
containing 2333 universities and 134330 student mobilities. Two universities are connected in the network, if there 
has been a student exchange among them. The number of Erasmus agreements a university posses are increasing in time. If one assumes a constant growth rate $\mu$ for this growth and an exponential increase in the number of Universities, the exponential degree distribution results from our model.  

The same exponential tail is observed for the Hungarian talent-supporting organizations collaboration network 
\cite{talent-net} in the in-degree distribution. We illustrate
the $\rho(n)$ dependence obtained from a logarithmically binned histogram on the right panel of 
Fig. \ref{EXPONET}. In 2014 there were recorded 1045 talent supporting organizations in Hungary, and with an online survey the central bureau mapped their collaboration network \cite{conference}. In the online questionary one had to indicate those networks with whom they had common activities. In total 4691 such directed links were revealed leading to a directed network. The number of registered talent supporting organizations is on the other hand growing fast as a function of time. Although there is no experimental evidence that this growth is exponential, one might assume such a growth for the years preceding this survey. One can speculate thus that the exponential nature of the in-degree distribution of this network can be explained with the same growth and dilution mechanism that has been discussed for the Erasmus collaboration network.

Preferential attachment dynamics in complex networks has also been considered.
It is natural to denote the number of connections to a given node by $n$ and then to assume 
a {\em linear preference} principle for adding
the next node and its connections. In this scenario, in the growth phase of networks, the transition rate going from
$n$ to $n+1$ connections is a linear function: $\mu_n=\sigma(n+b)$. Here very often $b=1$ is considered.
This picture has been extended to several phenomena showing network-like behavior, not only to physically
linked networks. E.g. cluster size growth shows a very similar dynamics \cite{KerteszKullman}.

It is seldom paid attention, however, to the resetting rate. 
This may stem either from a dilution of the sampling space or from real resetting processes,
when nodes disappear from the network.
We postulate here $\gamma_n=\gamma$ as a constant rate for the long jump from having $n$ connections to none.
When random failure of nodes happens ''democratically'', then the
rate $\gamma_n$ indeed must be independent of $n$. Popular nodes and unimportant nodes would be diminished
by the same rate in this scenario. This is in contrast to conscious attacks against a network \cite{BARABASIATTACK},
where nodes with many connections are preferred targets.

In the following discussion we restrict ourselves to the large $n$ limit,
investigating continuous distributions.
The linear preference rate, $\mu(x)=\sigma(x+b)$, and a constant resetting rate, $\gamma(x)=\gamma$, 
determine the Tsallis--Pareto form for the stationary distribution:
\be
 {\cal Q}(x) \sim \left( 1 + x/b \right)^{-1-\gamma/\sigma}.
\ee{TPNETWORK}
%Such stationary distributions emerge in all cases when $\mu(x)/\gamma(x) \sim x$ is linearly rising.
Indeed in Ref.\cite{TSTHUR} this degree distribution, called $q$-exponential, fits simulation data
nicely. In \cite{BarabasiREVMOD}
one finds many more examples for degree distributions exhibiting power-law type tails.  The linear preference in the attachment rate has been measured 
on the dynamics of internet topology, scientific collaboration and 
movie actor collaboration networks \cite{BARANEDAYEONG}. The authors found a preference rate
nearly linear: $\mu(x) \sim x^{1.05\pm 0.1}$ for the internet dynamics and 
$\mu(x) \sim x^{0.95 \pm 0.1}$ for citation networks.
On the other hand the scientific co-authorship and the movie actor collaboration
network shows a sublinear behavior, $\mu(x)\sim x^{0.80\pm 0.1}$.

\begin{figure}[hbt]
\begin{center}
\includegraphics[width=1.0\textwidth]{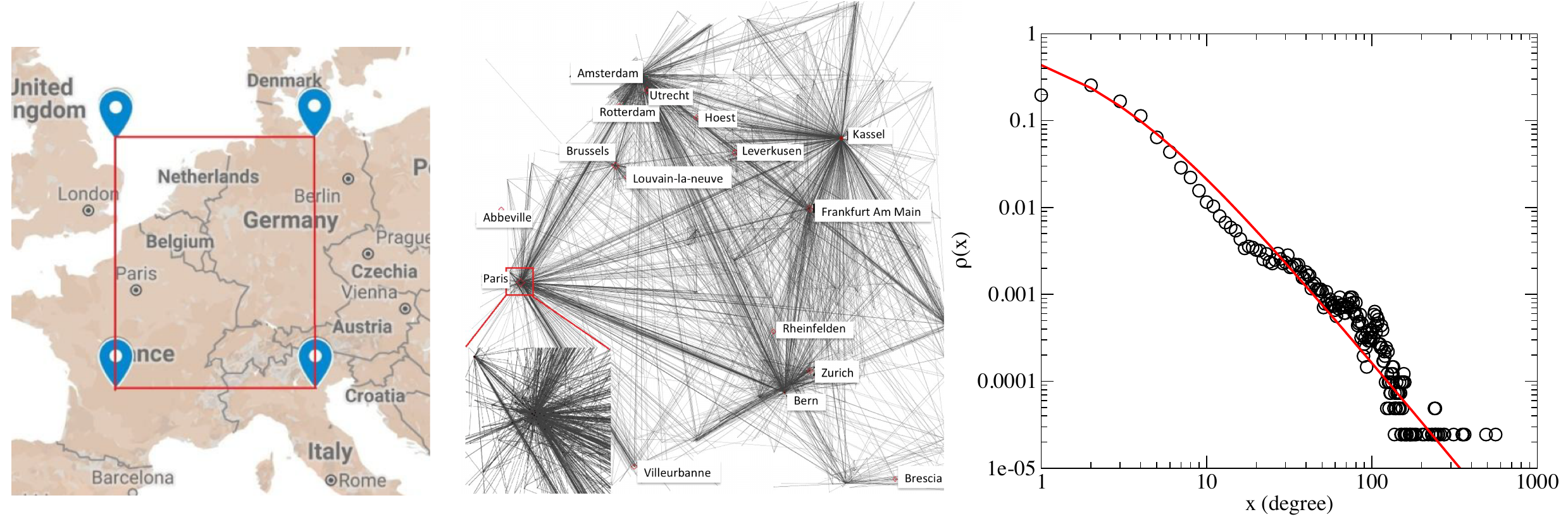}
\end{center}
\caption{\label{new}
 Results for the Internet wiring topology on router level. The left panel illustrates the mapped area of Europe. The panel in the middle sketches the unveiled wiring topology. The right panel shows the derived degree distribution (empty circles), with a Tsallis-Pareto fit [equation (\ref{TPNETWORK})] using $b=2.3$ and $\gamma/\sigma=1.3$. Figures replotted from \cite{MOUNIR} with the permission of the author.}
\end{figure}

To exemplify the presence of a scale-free tail in systems where the dynamics is governed by a linear preferential 
growth, we consider the topology of the Internet on router level and use the results obtained by Mounir Afifi in his master
dissertation \cite{MOUNIR}. He used the data provided  by the Centre for Applied Internet Data Analysis (CAIDA) \cite{CAIDA}.
The datasets contain "Traceroute" measurements conducted from CAIDA monitors, involved in the Archipelago project. Most of the data comes from monitors located in the Netherlands and Switzerland, which were supplemented with data from monitors in Spain, Germany, Finland, and Sweden. Using this data and locating the routers after their IP addresses we constructed a
map of the Internet topology in a restricted area of Europe (left panel in Figure \ref{new}). The topology of the Internet wiring is
illustrated in the middle panel of Figure \ref{new}. The degree distribution derived from this graph can be well fitted 
by the Tallis-Pareto distribution [eq. (\ref{TPNETWORK})] with $b=2.3$ and $\gamma/\sigma=1.3$. This is illustrated in the right panel of Figure \ref{new}.  	

We have briefly reviewed here the two most frequent degree distributions found in complex networks.
Definitely one may find even further fits or regressions under various constraints.
However, it shall be a deeper level of model making when one associates growth and resets
rates to explain the degree distributions found experimentally. 
Finally, we have to mention here that master equations were used for modelling small-world and scale-free network topologies
also in Refs. \cite{MASTERNET,WARINGNET,AGINGNET}

%%%%%%%%%%%%% HERE OCT 7

\subsection{Distribution of citations}

In a recent study concerning the statistics of citations to scientific
works and shares of Facebook posts \cite{PLOSONE2017} we investigated 
such a growth model. Denoting the number of
$n$ times cited papers or Facebook posts at time $t$ by $N_n(t)$, we allow for a preferential
rate $\mu_n$, for adding one new citation. Then the dynamics is described by the  chain 
\be
 \dot{N}_n \: = \: \mu_{n-1}N_{n-1} - \mu_n N_n
\ee{CITDYN}
for all bins $n \ge 1$, while the number of zero cited (fresh new or never cited) articles and posts
come with an increasing rate
\be
 \dot{N}_0 \: = \: \gamma N(t) - \mu_0 N_0,
\ee{CITZERORATE}
with $N(t)$ being the total number of papers published until that time
\be
 N(t) \: \equiv \: \sumi{n} N_n(t).
\ee{DEFTOTAL}
This leads to an exponential dilution of the background:
%It is easy to see that this total number grows with an exponential rate
\be
 \dot{N} \: = \: \sumi{n} \dot{N}_n \: = \: \gamma N,
\ee{TOTALRATE}
resulting in $N(t) \sim \eon{\gamma t}$. 
Several examples indicate that the total number of publications and Facebook posts
indeed grows exponentially \cite{PLOSONE2017}.
The corresponding equation for the fraction of $n$-times cited
posts, $P_n(t)\equiv N_n(t)/N$,  in this case is equivalent to eq.(\ref{PDOTUNIRESET}).

\begin{figure}
\begin{center}

\includegraphics[width=0.9\textwidth]{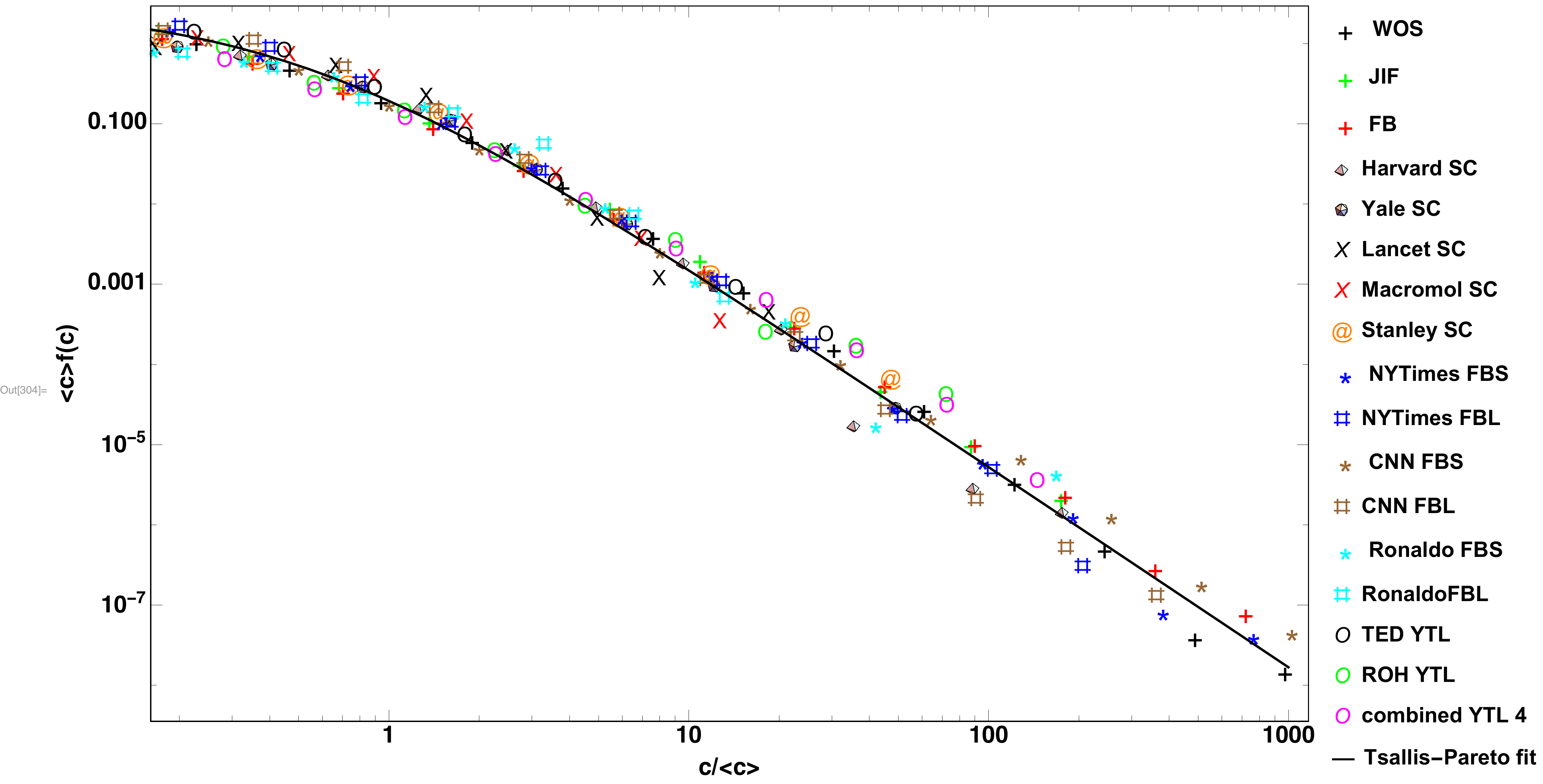}  

\end{center}
\caption{ \label{CITFIG}
 Rescaled distribution of citation numbers, $\exv{c} f(c)$, vs the ratio $c / \exv{c}$.
 WOS stands for the Web of Science, FB for the Facebook, JIF for Journal Impact Factor, 
 TED for the Ted Talks channel, and ROH for the Royal Opera House in London. Stanley is
 a well-known Professor of Physics, Ronaldo a soccer player, and NYTimes abbreviates 
 the newspaper New York Times. Lancet and Macromolecules are scientific journals.
 Harvard and Yale denote the scientific output of the corresponding universities in
 a decade.
 The type of citations are encoded as follows: SC for scientific citations, FBS for Facebook
 shares, FBL for Facebook likes, and YTL abbreviates you tube likes. The black curve
 represents the Tsallis--Pareto fit with a single parameter in its scaling form: 
 $a=2$.
}
\end{figure}

The total number of citations is given by the sum
\be
 C(t) \: \equiv \: \sumi{n} n \, N_n(t).
\ee{CTOTAL}
Its time derivative, using the above definitions, becomes
\be
 \dot{C} \: = \: \sumi{n} n \, \dot{N}_n \: = \: \sumi{n} \mu_n N_n.
\ee{DOTC}
Assuming again a linear preference rate 
\be
 \mu_n \: = \: \sigma ( n + b),
\ee{CITPREFRATE}
we get the following dynamics of the total number of citations:
\be
 \dot{C} \: = \: \sigma \left( C + b N \right).
\ee{DOTCMATE}
The solution is given by
\be
 C(t) \: = \: C_0 \eon{\sigma t} \, + \, \frac{\sigma b}{\gamma-\sigma} \, N_0 \, \left( \eon{\gamma t} - \eon{\sigma t} \right).
\ee{CSOLUTION}
The average citation per post behaves as:
\be
 m(t) \: = \: \frac{C(t)}{N(t)} \: = \: \frac{C_0}{N_0} \eon{(\sigma-\gamma)t} \, + \,
 \frac{\sigma b}{\gamma-\sigma} \, \left( 1 - \eon{(\sigma-\gamma)t} \right).
\ee{AVCIT}
In realistic cases, like PubMed, Web of Science database and some popular Facebook posts (NASA, New York Times, Ronaldo)
one has $\gamma > \sigma$ and the average citation per post tends towards a constant \cite{PLOSONE2017},
\be
 \lim_{t\to\infty}\limits m(t) \: = \: \frac{\sigma b}{\gamma-\sigma}.
\ee{AVCITLIMIT}
By all this, the stationary distribution $Q_n = \lim_{t\to\infty}\limits P_n(t)$ 
is a Waring distribution, in the continuous limit a Tsallis--Pareto distribution:
\be
 Q(x) \: = \: \frac{\gamma}{\sigma b} \, \left(1 + \frac{x}{b} \right)^{-1-\gamma/\sigma}.
\ee{TSPARWITHGAMSIG}
This distribution scales like
\be
 {\cal Q}(x) \: = \: \frac{1}{\exv{x}} \, f\left(\frac{x}{\exv{x}} \right)
\ee{TSALLISSCALING}
with $f(y) \: \equiv \: (a+1) \, \left(1+ay \right)^{-2-1/a}$ and $a=\sigma/(\gamma-\sigma)$, as the only fit parameter. 
For a finite $\exv{x}=ab$ one needs $a > 0$, i.e. $\gamma > \sigma$.
This scaling is demonstrated on various citation and Facebook share data in Fig.\ref{CITFIG}.
Technical details of obtaining these data are described in \cite{PLOSONE2017}, and here 
in Fig.\ref{CITFIG} we present an essentially wider survey of various data including scientific citations, 
Facebook shares and likes and You Tube likes as well.

%%%%%%%%%%%%%%%%%%%% FIG: CITFIG %%%%%%%%%%%%%%%%%%%%%%%%%%
%%%%%%%%%%%%%%%%%%%%%%%%%%%%%%%%%%%%%%%%%%%%%%%%%%%%%%%%%%%%

In closing this subsection we mention that master equations used in another perspective were involved already in the early 1984 for  explaining the citation dynamics \cite{MASTERCIT}.

\subsection{Hadron production}

A simple  mechanism will be considered for the hadron production 
process in high energy heavy-ion and elementary particle collisions.
In such collisions a high energy density state is formed, with a huge number of quarks and gluons.
It is treated either as an ensemble of partonic jets, inside which a quasi one-dimensional
gas of subhadronic partons are present, or a three-dimensional, high temperature
strongly coupled quark-gluon plasma. These systems freeze out at the end of a sudden
cooling process giving birth to hadrons.

The number of hadrons made, $n$, is called the multiplicity of a single collision event.
In millions of repeated collisions the distribution of this hadron number,
the multiplicity distribution, $P_n$, is evolving during the process.
We assume that the measured multiplicity distributions are close to the stationary
distribution, $Q_n$.

For these hadronization processes  we construct the following simple model:
Having already $n$ hadrons, a new hadron is created with the probability rate $\mu_n$,
and a collective re-melting into the prehadron stage happens with the rate $\gamma_n$.
We conjecture that there is a certain number of newly made hadrons, $\exv{n}$, for which the
re-melting does not occur.
Less hadrons than this number will likely to be created from a zero number state,
featuring formally negative $\gamma_n$ rates for $n < \exv{n}$. In this case the
yield is proportional to $P_n$, not with $P_0$.
More hadrons, $n > \exv{n}$, will be re-melted with a positive $\gamma_n$ rate in this scenario.
We work thus with the ansatz $\gamma_n=\sigma(n-\exv{n})$. 

Furthermore we assume
$\mu_n = \sigma (n/k+1)\exv{n}$, realizing the Matthew principle with linear
preference. Certainly, already at $n=0$, in the state with no hadrons, there is a probability rate
to create one, $\mu_0=\sigma\exv{n}$. Already having $n$ hadrons on the other hand helps this process; most of the
hadrons to be made are light bosons, mainly pions. 
The construction of the rates are harmonized for $\gamma_0+\mu_0 = 0$.
For all other $n$ values one has $\gamma_n+\mu_n > 0$.

Accepting all these assumptions, the obtained stationary distribution in our master equation framework 
is the negative binomial one,
\be
 Q_n \: = \: 
\binom{n+k-1}{n} \, 
\frac{(\exv{n}/k)^n}{(1+\exv{n}/k)^{k+n}}.
\ee{HADRONNBD}
The first moment turns out to be exactly $\exv{n}$.
This distribution is also generated by the $k$-th power of the geometrical series \cite{BiyajimaWilk},
\be
 (1-x)^{-k} \: = \: \sumi{n} \binom{n+k-1}{n} \, x^n,
\ee{NBDGENER}
as a generalization of the binomial formula for negative powers. 
The second scaled factorial moment,
\be
 F_2 \: \equiv \: \frac{\exv{n(n-1)}}{\exv{n}^2} \: = \: 1 + \frac{1}{k} \: > \: 1
\ee{NBDF2}
indicates that this distribution is super-Poissonian. In the $k\to\infty$ limit the
Poisson distribution emerges when keeping $\exv{n}$ finite.

Finally a remark on the negative values of certain $\gamma_n$-s: Is it fatal?
No, as long as $\mu_n+\gamma_n > 0$, the recursion solution \re{QRECURN} for $Q_n$ works fine.
In the above outlined hadronization scenario $\gamma_n+\mu_n = \sigma n(1+\exv{n}/k) > 0$, this
condition is fulfilled. The meaning of sometimes positive sometimes negative $\gamma_n$
resetting rates represents in the state space such a long jump process, which under
certain circumstances goes in the opposite direction. However, in this case the
dependence on the initial state occupation probability also turns into the same dependence
on the final state probability; this is a special, up to now scarcely investigated
skew symmetry principle for random processes not satisfying detailed balance condition.
Our proof for the decrease of the entropic divergence to the stationary distributions
is not directly applicable to this system. We postpone this problem for a future work.

\begin{figure}
\begin{center}

\includegraphics[width=0.75\textwidth]{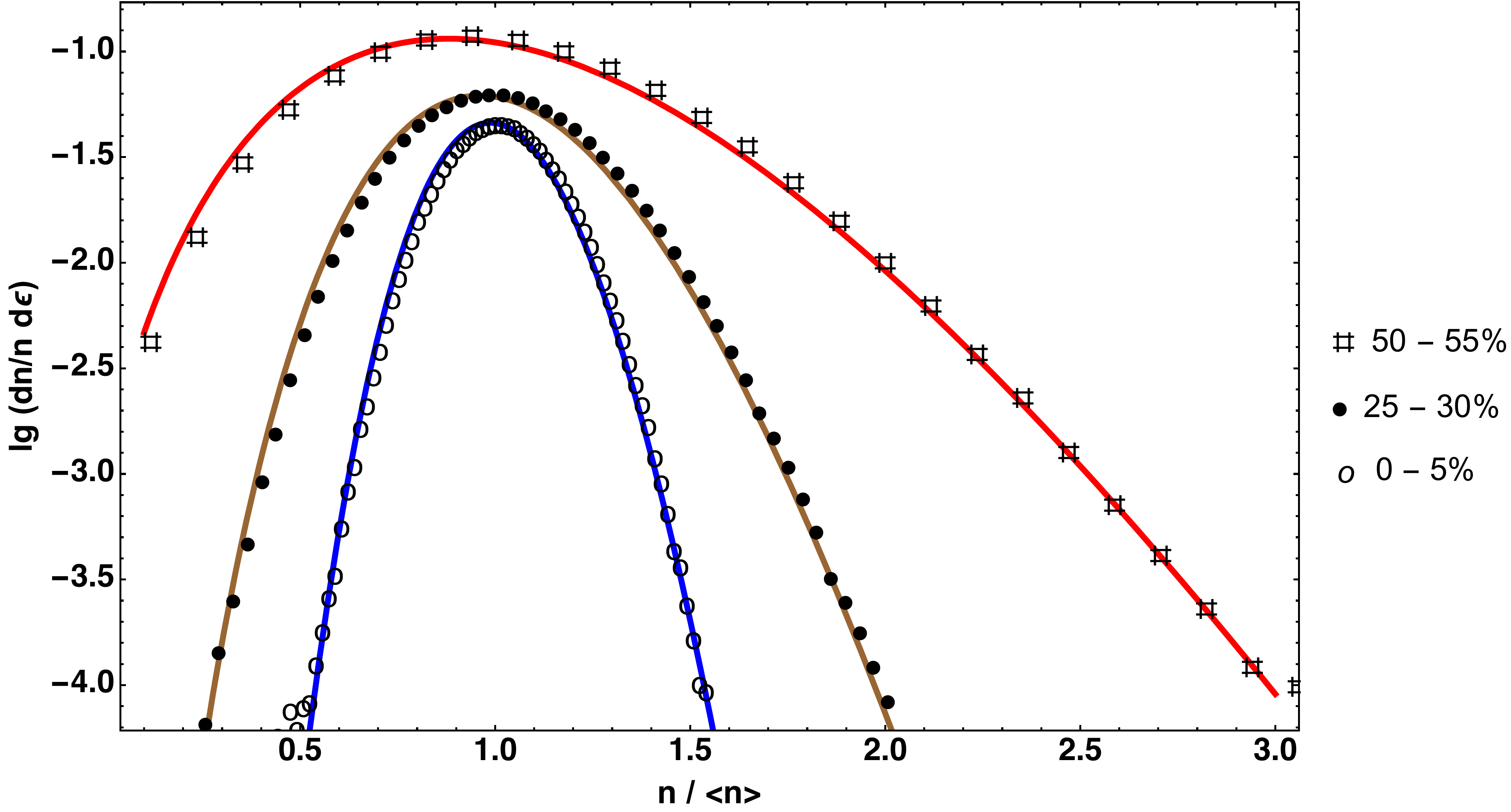}

\end{center}
\caption{ \label{PHENFIG}
 Total charged hadron multiplicity distributions from  Au+Au collisions  at $200$ GeV
 center of mass energy per incoming nucleon, plotted for different centrality classes: 
 for (a) $0-5$\%, for (b) $25-31$\% and for (c) $55-60$\%, form the bottom to the top.
 The NBD best fit parameters are as follows: (a) $\exv{n}=61$, $k=255$; (b) 
 $\exv{n}= 27.4$, $k=50 $; (c) $\exv{n}=8.5$, $k = 17$.
}
\end{figure}

Experimental findings for hadron multiplicity distributions in high energy experiments indeed
show results well approximated by the negative binomial (NBD) distribution.
A demonstrative example is given by results of the PHENIX experiment at the
Relativistic Heavy Ion Collider operating at Brookhaven \cite{PHENIX}.
Based on publicly available data we have replotted some of the multiplicity distributions
for the reaction AuAu at $200$ GeV on Fig.\ref{PHENFIG}. On the horizontal axis
the ratio $n/\exv{n}$ is given and on the vertical axis we have the measured normalized jet data, related to the   
$Q_n$ distributions. The NBD distribution (\ref{HADRONNBD}) with parameters given in the figure caption 
offers a fair fit. 

The NBD multiplicity distribution also leads to an approximate Tsallis--Pareto distribution
in the single particle kinetic energy. Considering the probability of having $n$ particles,
$P_n$, sharing the total energy $E$ we have two, physically different cases, described by  mathematically
equivalent formulas. In a high-energy jet of partons, moving only in a very narrow
cone relative to the leading parton, the phase space is exactly
$n$-dimensional with the total relativistic kinetic energy (with $c$ being the speed of light):
\be
 E \: = \: c \sum_{j=1}^n\limits |p_j|.
\ee{JETKIN}
At low energy the typical measurement for the hadronic fireball is a two-dimensional
transverse momentum distribution. The ensemble of $n$ particles live in a $2n$-dimensional kinetic
phase space, with the total non-relativistic energy for slow particles
\be
 E \: = \: \frac{1}{2m} \sum_{j=1}^{2n}\limits p_j^2. 
\ee{TWODIMKIN}
The third dimension is suppressed in the transverse momentum spectra by the narrow rapidity selection.

In both cases the probability density of finding a given energy value, $E$, 
is proportional to the occupied phase space volume, $\Omega(E)$.
This is the simplest assumption underlying the Boltzmannian statistics.
The fixed total energy shell phase space volume can be derived from the volume of the $N$-ball,
constructed according to the kinetic energy formula. One realizes that
\be
 \Omega_N(E) \: = \: \int\!\delta(E' - E ) \, d\Gamma_N \: = \:
 \frac{1}{dE/dR} \, \int\!\delta(R'-R) \, d\Gamma_N
 \: = \: \frac{1}{dE/dR} \, \pt{}{R} V_N(R) \: = \: \pt{}{E} \, V_N(R(E)).
\ee{SHELLVOLUM}
The phase-space volume for an $N$-ball in an $L_p$ norm is given by:
\be
 V_N^{(p)}\left( \Big( \sum_i\limits |x_i|^p\Big)^{1/p} \le R(E)  \right) \: = \: \frac{\Gamma(1+1/p)^N}{\Gamma(1+N/p)} \, (2R)^N.
\ee{NBALLLP}
The $p=1$ case with $R(E)=E/c$ describes strongly relativistic jet particles, while the $p=2$
case with $R(E)=\sqrt{2mE}$ non-relativistic, massive particles.
The corresponding hypershell volumes, $\Omega_N^{(p)}(E) = dV_N^{(p)}(E)/dE$, are given as
\be
  \Omega_n^{(1)}(E) \: = \: \pt{}{E} \frac{(2E/c)^n}{n!} \: = \: \left(\frac{2}{c}\right)^n \, \frac{E^{n-1}}{(n-1)!},
\ee{JETHYPERVOL}
for the jets with $N=n$ kinetic degrees of freedom and 
\be
 \Omega_{2n}^{(2)}(E) \: = \: \pt{}{E} \frac{(2m\pi E)^n}{n!} \: = \: (2m\pi)^n \, \frac{E^{n-1}}{(n-1)!}
\ee{TWODIMHYPERVOL}
for the massive particles moving in two dimensions with $N=2n$.

The {\em single particle energy spectra} reflect the hypershell volume ratio
\be
 r_{N,g}^{(p)} \: \equiv \: \frac{\Omega_{g}^{(p)}(\epsilon) \; \Omega_{N-g}^{(p)}(E-\epsilon)}{\Omega_N^{(p)}(E)},
\ee{PHVOLRAT}
with $g=N/n$ single particle degrees of freedom.
This ratio coincides for $N=n$ and $p=1$ (quasi one-dimensional jet with $n$ particles) 
with the value at $N=2n$ with $p=2$ (two-dimensional non-relativistic gas with $2n$ momentum components):
\be
 r_{n,1}^{(1)} \: = \: r_{2n,2}^{(2)}  \: = \: \frac{n-1}{E} \, \left(1-\frac{\epsilon}{E} \right)^{n-2}.
\ee{RATIO}
This formula is nonzero only for $n\ge 2$ (the minimum number of particles to share energy is two),
and terminates for $\epsilon > E$, which is the maximal energy for the selected particle.
We note that by construction
\be
 \int_0^E\limits r_{N,g}^{(p)}(\epsilon,E) \, d\epsilon \: = \: 1.
\ee{NORMOFRATIO}
Obtaining single particle energy spectra from high-energy experiments usually overlays results for
different hadron multiplicities, $n$. Having a probability distribution for the newly produced hadrons, $P_n$,
(please note that also minimum two particles are necessary to initiate a collision), one presents single
particle spectra as
\be
 \frac{1}{{\cal N}} \pt{{\cal N}}{\epsilon} \: = \: \sum_{n=2}^\infty\limits r^{(p)}_{ng,g}(\epsilon,E) \, P_{n-2}.
\ee{SPSPEC}
The jet and massive two-dimensional gas can be treated with the common formula, eq.(\ref{RATIO}) here.
This also might be the reason why the low-momentum and high-momentum parts of observed $p_T$-spectra
can be covered by a single fit.

\begin{figure}
\begin{center}

\includegraphics[width=1.00\textwidth]{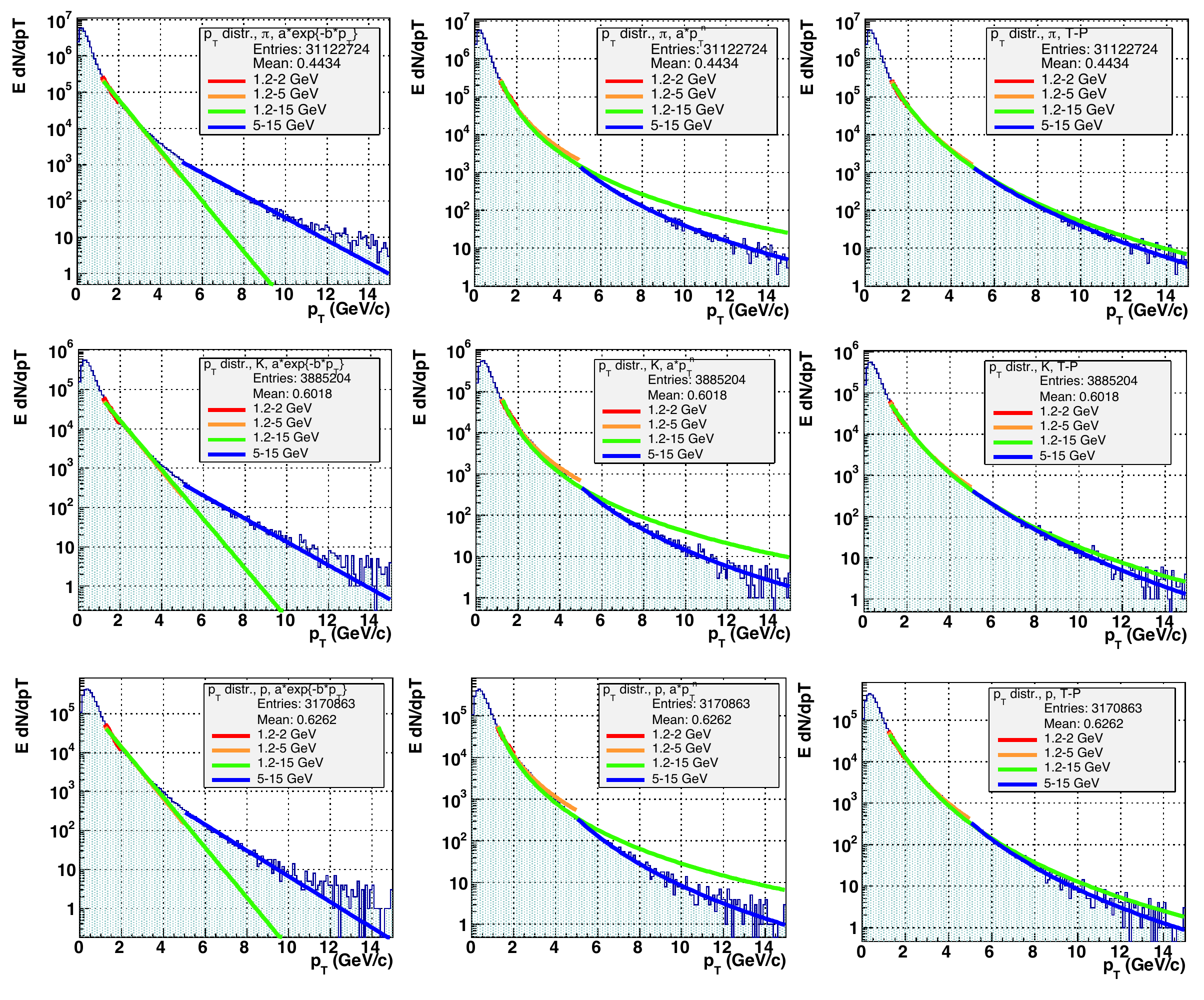}

\end{center}
\caption{\label{FIG:SPECTRA}
 Transverse momentum spectra of different hadrons (pions, kaons and protons from top down)
 and three different fits to them: exponential (left column), power-law (middle column) and
 Tsallis--Pareto distributions (right column). Figures from the master thesis of G\'abor
 B\'{\i}r\'o, with the permission of the author.
}

\end{figure}

Using the negative binomial distribution (\ref{HADRONNBD}) one obtains
\be
 \frac{1}{{\cal N}} \pt{{\cal N}}{\epsilon} \: = \:
 \frac{1}{E} \left(1+\frac{\exv{n}}{k} \, \frac{\epsilon}{E} \right)^{-k-1} \, 
 \left[1 + \frac{\exv{n}}{k}\, \frac{\epsilon}{E} \, + \, \exv{n} \left(1-\frac{\epsilon}{E} \right) \right].
\ee{SINGLESPECFROMNBD}
It has a finite value both at $\epsilon=0$ and at $\epsilon=E$. The latter is due to the fact, that the
hypershell volume ratio for making no new hadrons is one, which has to be interpreted as the Heaviside
theta function $\Theta(E-\epsilon)$. For small energies, $\epsilon \ll E$, up to a tenth of $E$,
the above result is well-approximated by a Tsallis--Pareto distribution:
\be
 \frac{1}{{\cal N}} \pt{{\cal N}}{\epsilon} \: \approx \:
 \frac{\exv{n}+1}{E} \left( 1 + \frac{\exv{n}}{k} \, \frac{\epsilon}{E} \right)^{-k-1}.
\ee{APPROXTSALLISNBD}
Hadronic transverse momentum ($p_T$) spectra are indeed best fitted by Tsallis--Pareto
distributions. Fig.\ref{FIG:SPECTRA} shows such experimental findings and different fits,
carried out by G\'abor B\'{\i}r\'o in his MSc thesis at the E\"otv\"os University in 2016.
The upper row is for pions, the middle row for kaons and the lower row for protons, particles
with different rest mass. Figures in the left column stand for the Boltzmann exponential fit,
in the middle column for a pure power-law fit, and finally in the right column for the
Tsallis--Pareto fits. 
The different colored thick curves represent fits to different $p_T$ ranges, as given in the legends. 
It is obvious that only the Tsallis--Pareto form is able to cover both
higher and lower $p_T$ values in the experimental results, while the others fail for the values outside
the respective fit ranges \cite{BIROGABOR}.

\subsection{Income distribution}

Another good example is the income distribution in societies.  As a general rule one observes in the density function two well distinguishable regimes \cite{HIGHINCOMEREVIEW}.  One regime is the high income region (usually the upper 2\% of the population),  where a clear power-law like trend is observable. On the other hand in the low and medium income region (98\% of the population) the income distribution can be well fitted by a gamma distribution. 

Let us consider the top income region (upper 2 \% of the population) first, from the perspective of our model.
Here the increase in the income is always a given per cent, not a given amount, so in our master-equation approach 
a purely linear increase rate will be considered: $\mu(x)=\sigma x$. A constant resetting rate can be assumed by a retirement/death process $\gamma(x)=\gamma$.  
The above assumptions leads to a  power-law tail in the distribution; in fact the original Pareto-law.
Such distributions are observed, only in the high-end tail of the income distribution
(top 2\% of the households) \cite{HIGHINCOMEREVIEW}.
The Pareto-exponent $\alpha$ obtained in observations is robust. It does not change over decades
of years, although the actors in the society and the persons in the top income category are continuously
changing \cite{DERZSY}. The Pareto exponent, in our model is given as $\alpha=1+\gamma/\sigma$, where the $\sigma$
coefficient in the linear preference law, $\mu(x)=\sigma x$, seemingly follows the
resetting rate $\gamma$.  It is interesting to note here that Pareto's law has been observed also in historical wealth data.
Beginning with ancient societies \cite{EGYPT}, through classical and medieval ages \cite{MEDHUN}
and also the beginning of capitalism \cite{PARETO}. Income distributions mapped nowadays from more precise,
electronically available data confirms it's generality \cite{DERZSY}. 

For describing the income distributions in the middle and low classes  
(98\% of the society), one has to consider a different choice for the  $\gamma(x)$ rates. 
We will consider the same linear form for the growth rate $\mu(x)$ for persons that are already in the system at a given time moment. Under such an ansatz the resetting rate $\gamma(x)$ is however more complicated. As one reaches a higher income, presumably gets older and
thus the resetting rate has to increase as well. This is very different from the very high income region, where we assumed the same resetting rate for all income categories. One also has to take into account that employees that had no income can appear 
in an income category (beginners), but with a bigger probability in the income-categories below the average.     
 We consider thus a smart resetting rate which can become also negative, best balanced at an intermediate income class and increasing with high income. 
The most simplest linear choice for the $\gamma$ rates that incorporates all the above discussed effects and which may go
also to negative would be $\gamma(x)=\sigma(ax-c)$.  

\begin{figure}[hbt]
\begin{center}
 \includegraphics[width=0.5\textwidth]{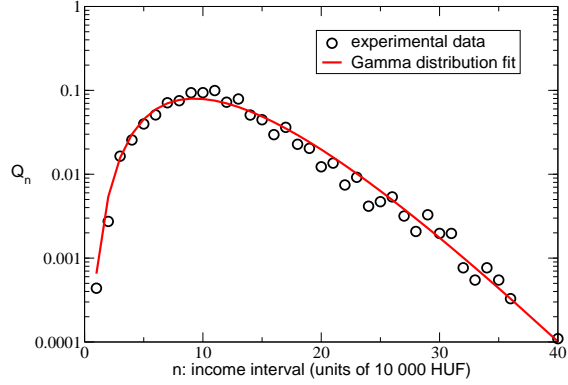}
\end{center}
\caption{\label{HUNINCOME}
 Hungarian income distribution in 2014 according to a non-official survey of KSH in thousand Hungarian Forint
units. Dots are binned data and the continuous line is a gamma distribution fit (\ref{INCOMEGAMMA}) with $c=4.6$ and $a=0.3883$}
\end{figure}

The above assumptions lead to a gamma distribution at stationarity:
\be
 {\cal Q}(x) \: = \: \frac{a^c}{\Gamma(c)} \, x^{c-1} \, \eon{-ax}.
\ee{INCOMEGAMMA}
Here $\Gamma(c)$ denotes Euler's gamma function with the property $\Gamma(c+1)=c\Gamma(c)$.
Certainly, this $\gamma(x)$ is also negative for $x < c/a = \exv{x}$. 

As a demonstration we present the result of a partial net income survey made by the
Central Statistics Bureau (KSH) in Hungary in 2014. The data sample, binned by 10,000 HUF bins,
is nicely approximated by a gamma distribution with $c=4.6$ and $a=0.3883$, resulting
in the average income of $\exv{x} =c/a= 118,465$ HUF. The maximum of the fitted curve is at $x_m=(c-1)/a=92,711$ HUF,
this was the net income most people had.

%\subsection{Biodiversity and Settlement Size Distribution}
\subsection{Biodiversity}

A major challenge in ecology is to understand the abundance distribution in
communities of neutral species \cite{BIOHUBBLE}. Given an ensemble of species which do
not compete with each other, except sharing a common ecological niche, the abundance
distribution describes the PDF, $Q_n$, for having $n$ individuals in one of the species.

The most popular and well-know fit to the observed distribution used by ecologists
is the famous Fisher's log series which suggests for the number of species with $n$ individuals the value \cite{BIOPDF}:
\be
 S_n \: = \: \alpha \, \frac{a^n}{n}
\ee{BIOBS}
where $\alpha$ and $a$ can be determined knowing the total number of individuals,
\be
 N \: \equiv \: \sum_{n=1}^{\infty}\limits n S_n \: = \: \alpha \, \frac{a}{1-a},
\ee{NUMINDIV} 
and the number of species in the studied territory,
\be
S \: \equiv \: \sum_{n=1}^{\infty}\limits S_n \: = \: - \alpha \, \ln(1-a).
\ee{NUMSPECI}
This leads to 
\be
S \: = \: \alpha \, \ln \left(1+\frac{N}{\alpha} \right)
\qquad {\rm with} \qquad 
a \: = \: \frac{N}{\alpha+N}.
\ee{BIOBSNORM}

Nowadays exhaustive data sets are available for tree communities on large territories.
One of the most known is the Barro Colorado Island Tropical Tree Census (BCI) \cite{BCI},
administrated by the Smithsonian Tropical Research Institute in the U.S.
In this census more than 240,000 stems and over 300 tree and shrub species are
accurately mapped. Data are publicly available on request \cite{BCI}. 

\begin{figure}
\begin{center}

\includegraphics[width=0.85\textwidth]{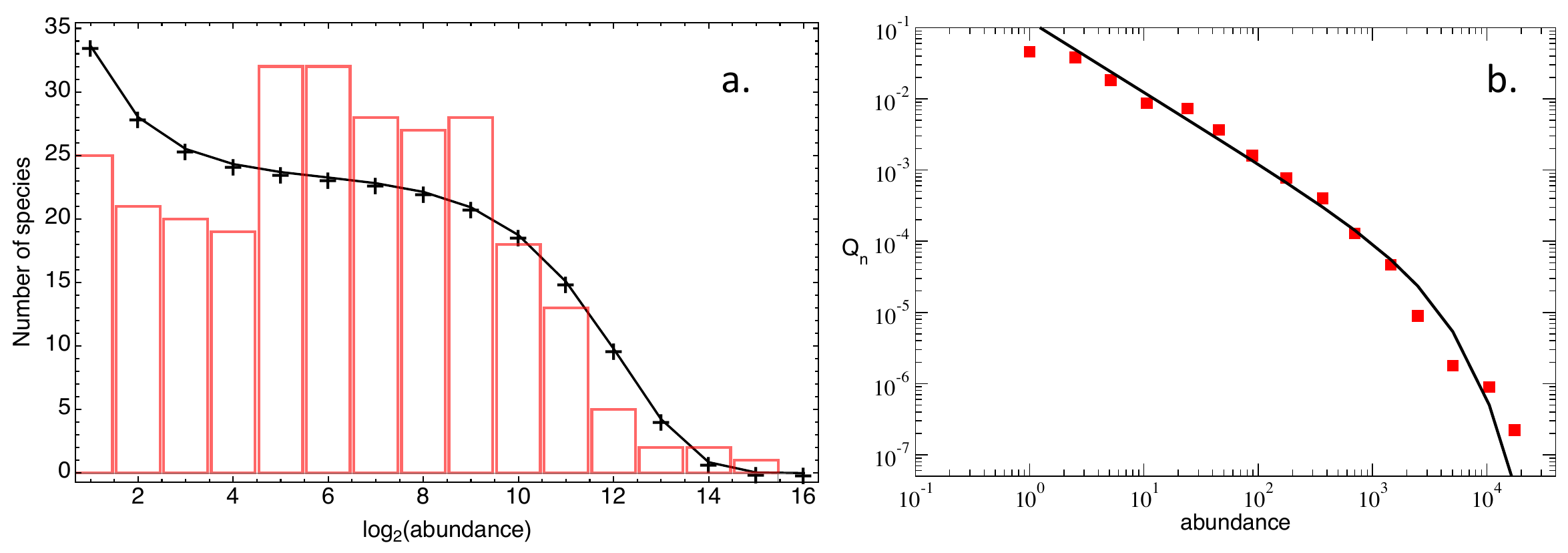}

\caption{\label{WOODS}
Tree species abundance distribution for the 1995 BCI data. Fig. a presents the Pearson plot ($W_k$ with $k=\log_2 n$), Fig. b
the PDF ($Q_n$). Continuous lines in both parts stand for the  Fisher log series fits.
}

\end{center}
\end{figure}

As an example for the appropriateness of the Fisher log series fit \cite{NEDAJTB}, on  Fig.\ref{WOODS}
we present abundance distribution from the BCI Tropical Tree Census in 1995 
determined from a 25 ha territory containing $N=112 543$ trees belonging to a total number of 
$S=273$ species. Two different plots are shown. One is the Preston-type plot (Figure a.), where we plot the number of species found in abundance intervals of consecutively doubling lengths. Preston’s method of plotting is motivated by the
fact that abundances can vary over several orders of magnitude and there are far fewer abundant species than rare 
ones. Using fixed-length abundance intervals would result in large statistical
fluctuations at the tail of the curve. In Fig.\ref{WOODS}a  we indicated by bars the number of species found in such increasing abundance intervals and with black symbols the number predicted by the Fishers log-series
distribution: 
\begin{equation}
W_k=\sum_{n=2^k}^{2^{k+1}-1} S_n.
\end{equation}
 Please note that the value of $\alpha$ and $a$ are not fitted, they are computed from the values of $S$ and $N$ 
based on equations (\ref{BIOBSNORM}). In the second plot (Figure b.) we have shown the PDF of the distribution,
i.e. $Q_n=S_n/S$ computed from the BCI data and the one calculated from equation (\ref{BIOBS}). 

The log series abundance distributions can easily be reproduced in the framework of our model, 
offering a possible explanation for the mechanism behind forming such a distribution. 
Let us assume a growth phenomena with resetting, and start from an empty territory.
We assume a pure linear preference rate for the
local growth factor, $\mu_n=\sigma n$. Then apply the recursion rule for the stationary
distribution (\ref{QRECURN}) using the log series prediction given eq.(\ref{BIOBS}). We obtain
\be
 \sigma (n-1) \, \alpha \frac{a^{n-1}}{n-1} \: = \: (\sigma n + \gamma_n)  \, \alpha \frac{a^n}{n}.
\ee{BIORECUR}
The solution for the resetting rate, $\gamma_n$, is also linear:
\be
 \gamma_n \: = \: n \sigma \left( \frac{1}{a} - 1 \right).
\ee{BIOGAMMA}
Since $a<1$ for $Q_n$ being normalizable, this reset rate is always positive.
We interpret it as a rate of a total destruction of all individuals in the species,
as an extinction rate (describing the frequency of long jumps from $n$ to zero in a short time step).
Its linear dependence on the number $n$ suggests that the extinction rate, increases with the size of the species
suggesting that larger species are more vulnerable. 

The approach presented here is on the mean field level without caring for spatiality.
More advanced models also consider the distribution of the individuals and species in space and their 
correlations \cite{NEDA2}. 

\subsection{Settlement size distribution}

\begin{figure}
\begin{center}

\includegraphics[width=0.47\textwidth]{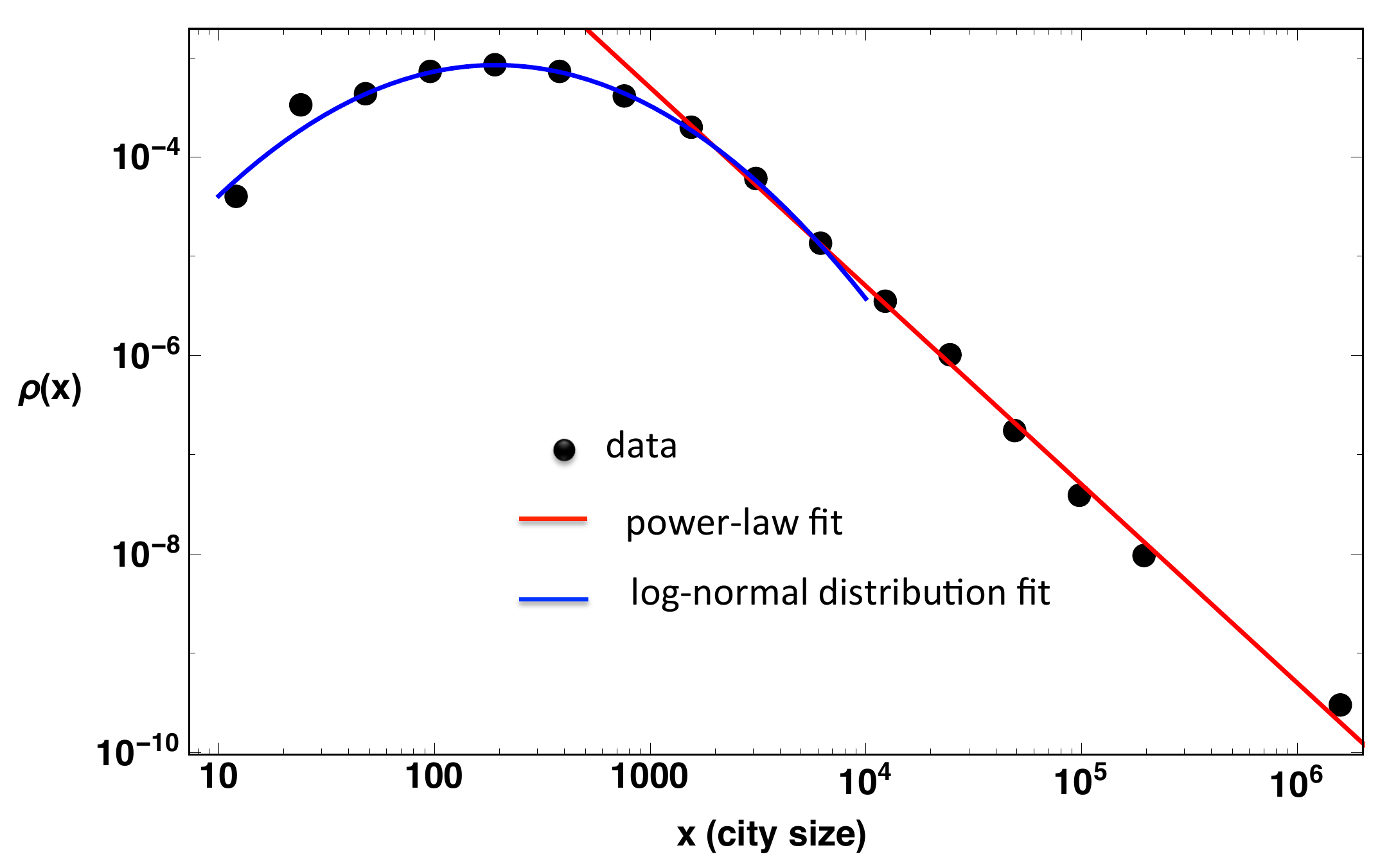} ~\hfill
\includegraphics[width=0.47\textwidth]{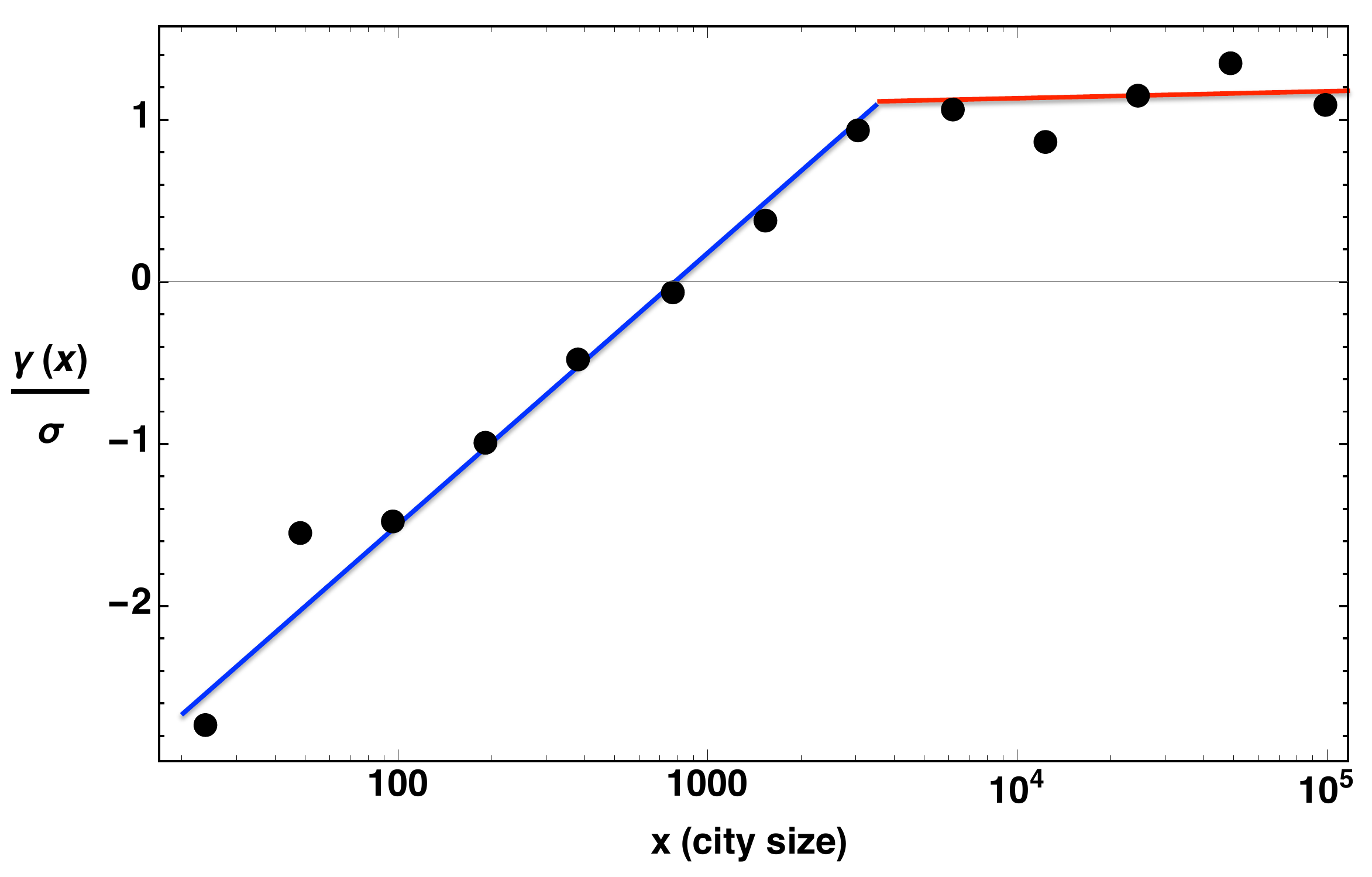}

\end{center}

\caption{\label{HUNCITY}
(Left panel)  Density function for the Hungarian settlement population size distribution on a log-log plot. The log-bined data
is presented by black dots, the power-law trend with exponent $\alpha = -2$ is illustrated by the red line, the log-normal fit to the
small and middle size settlement region is illustrated by the blue line ($\sigma = 1.2$ and $m = 6.7$).  (Right panel)
 The $\gamma(x)/\sigma$ value numerically derived from equation (\ref{GAMFROMOTHERS}).
}
\end{figure}

A further interesting application field is the study of settlement size distribution.
Data from different countries (US, France, Japan, China, India) follow similar trends.
Most fits show a log-normal distribution around the maximum frequency (middle sized cities),
and at the same time a high-end distribution tail following Pareto's law \cite{DECKER}. 
We illustrate this trend for the settlement size distribution in Hungary. On the left panel of Figure 
\ref{HUNCITY} we show the density function for the settlement-size distribution of all 3154 settlements in Hungary. 
The black dots represent the computed density function using a logarithmic binning method.
The blue curve shows a log-normal fit
\be 
\rho(x)=\frac{1}{x\sigma \sqrt{2\pi}} \exp{\left( -\frac{(\ln(x)-m)^2}{2\sigma^2} \right)}
\ee{LOGNOMAL}
with $\sigma=1.2$ and $m=6.7$. The red curves illustrate a power-law trend $\rho(x)\propto x^{\alpha}$ with 
$\alpha=-2$.   
The above observations turn our attention to the possibility that for large settlements
a different dynamics is realized than for small and middle sized ones. This is somehow similar with the case 
discussed in the subsection devoted to the income distribution. 
In order to elaborate on  this guess, we  use again the framework of our model and take a different approach now.
Having data on ${\cal Q}(x)$ and assuming a certain expression for 
the local rate, $\mu(x)$, we can express the nonlocal rate, $\gamma(x)$,
based on the stationary solution to the evolution equation (\ref{MASTERFLOW}). In such a view 
one can determine  the unknown model parameters necessary for reproducing the observed distributions. 
More specifically, the unknown $\gamma(x)$ will be given by the continuous version of eq.(\ref{QRECURN}):
\be
 \gamma(x) \: = \: - \frac{1}{{\cal Q}(x)} \, \pd{}{x} \left( \mu(x){\cal Q}(x) \right).
\ee{CONTQRECUR}
This result allows also negative values for $\gamma(x)$. These can be interpreted as
nonlocal tranistion rates from the ground state directly to a finite $x$ size.
This situation reminds us to the situation encountered in the discussion of income
distribution and particle multiplicities in high energy experiments.

Considering in particular a linear preference rate in the local growth rate, $\mu(x)=\sigma x$,
we obtain
\be
 \frac{1}{\sigma} \, \gamma(x) \: = \: -1 \, - \, \pd{\ln {\cal Q}(x)}{\ln x}.
\ee{GAMFROMOTHERS}
In particular for a power-law tail, ${\cal Q}(x) \sim x^{-\alpha}$ one obtains
a constant nonlocal rate, $\gamma(x)=\sigma(-1+\alpha)$. It is positive for $\alpha > 1$,
which at the same time is the condition for the normalizability of ${\cal Q}(x)$.
The log-normal distribution on the other hand, 
\be
{\cal Q}(x) \: \sim \: \frac{1}{x} \, \exp \, \left[\, - \frac{a}{2} \, \ln^2\left(\frac{x}{b}\right) \, \right],
%{\cal Q}(x) \sim \frac{1}{x} \eon{ {\large - \frac{a}{2} \, \ln^2\left(\frac{x}{b}\right)} },
\ee{CITYLOGNORMAL}
assumes:
\be
 \gamma(x) \: = \: a \ln \frac{x}{b}.
\ee{LOGNORMALGAMMARATE}
To illustrate the picture outlined above, in the right panel  
of Fig.\ref{HUNCITY} we show the value of $\gamma(x)$ numerically derived from equation (\ref{GAMFROMOTHERS}) 
plotted on normal-log scale. Indeed the shape of $\gamma(x)$ on the normal-log scale
justify both assumptions: a linear rise at small and middle sized settlements and a saturation
to a constant value at large sizes.  One can observe on this picture that the transition between the 
two regimes correspond to the one sketched in the left panel of Fig. \ref{HUNCITY}, 
namely the region where a log-normal fit is appropriate and the region where the power-law trend is observable.

If one accepts the assumptions inherent in the unidirectional growth and reset model,
the $\gamma(x)$ versus $\ln x $ plot in this case helps to distinguish distribution
models more efficiently  than  the familiar ${\cal Q}(x)$ or cumulated plots.
The data presented for Hungary is in agreement with findings in other countries
(US \cite{USCITY}, India \cite{INDCITY}, China \cite{CHINCITY}, OECD countries \cite{OECD}).

%%%%%%%%%%%%%%% CONCLUSION %%%%%%%%%%%%

\section{Conclusion}

In conclusion we have reviewed a model for unidirectional growth
augmented by rare resets to the ground state with great potential 
of applications to complex systems. We have considered fundamental questions,
and presented a new result about the statibility of stationary PDF-s. The proof is based
on a careful analysis of the notion of entropic divergence (by some called entropic distance,
although its definition is usually not symmetric in the two distributions as arguments).
We have found that for a global approach towards the stationary PDF, as far as it exists,
in all linear and nonlinear master equation approaches {\em it is sufficient} that
the definition of entropic divergence is based on a core function with definite convexity.
This property holds for any positive transition rate from the state $|m\rangle$ to $|n\rangle$
in general as long as the PDF dependent factor in all terms involves the initial state
of the microtransition only, in form of a general positive function, $a(P_m)$. 
Then we investigated factorizing dependencies on the initial and final state occupation probabilities in such microtransitions, 
$w_{nm}a(P_m)b(P_n)$. 
We have found that only rates satisfying the {\em detailed balance}
condition ensure the steady decrease of the entropic divergence to the stationary PDF.

During this proof a surprising consequence arose. Using the traditional
kernel function for defining the entropic divergence, $\sigma(\xi)=-\ln\xi$, 
in case of a nonlinear power dependence on the initial state occupation in the elementary transitions,
$a(P_m)\sim P_m^q$, the resulting  entropic divergence to the stationary
PDF becomes non-Boltzmannian. This result is a beautiful generalization  of the
Kullback--Leibler entropic divergence.
Moreover, seeking a relative entropy interpretation
for this result, one realizes that the entropic divergence from the uniform
distribution to the stationary PDF is proportional to the difference of Tsallis--entropies,
and not to the R\'enyi ones (cf. eqs.(\ref{TSALENTROP},\ref{URENYIDIV})).

Beyond these general investigations we have concentrated to a particular class of
stochastic models, describing unidirectional growth and resetting processes. 
This simple approach reduces the possible
micro-transitions to a local one, from $|n\rangle$ to $|n+1\rangle$ with the rate
$\mu_n$, and assumes a resetting from any state $|n\rangle$ to the ground state $|0\rangle$
with a rate $\gamma_n$. In spite the compactness and austerity of its formulation
(only with two dynamical concepts) this model performs surprisingly abundant. 
First, it helps to understand theoretical questions:
i) the emergence of stationary distributions and a soliton-like approach to them,
ii) the validity of a formula between the elementary local and nonlocal rates through the stationary distribution,
reminding very much to the fluctuation--dissipation theorem,
iii) the continuous  limit of discrete step-like processes ($n \rightarrow x$),
and iv) the reproduction of a number of well known and widely used distributions as stationary PDF-s
based on simple assumptions on the rates $\mu(x)$ and $\gamma(x)$.
Our presentation offers a straightforward alternative to the historical derivations. 
Second, it proves to be useful for a long list of applications:
i) complex network degree distributions,
ii) citation distribution of scientific papers and Facebook post shares,
iii) hadron production in high energy experiments,
iv) income and wealth distribution,
v) biodiversity
vi) city size distribution, and possibly further statistical phenomena, not discussed here.

In most applications a linear preference was assumed in the local growth rate, $\mu(x) \sim x$,
and a constancy in the resetting rate, $\gamma(x) = \gamma$, with a few notable exceptions.
From the linear local rate -- constant resetting rate dynamics a Tsallis--Pareto PDF emerges.
In the case when both type of rates are constant, the stationary PDF is exponential,
and finally with the resetting rate growing as the logarithm, $\gamma(x) \sim \ln x$,
and the local growth rate linearly, a log-normal distribution is established. Certainly,
further stationary distributions can belong to various different assumptions about
these rates. We would like to call the attention also to the reverse method: having
information about the final distribution, ${\cal Q}(x)$, and one of the rates, the other one
can be obtained by the statistical fluctuation--dissipation relation. This opens a novel
view to the analysis and interpretation of numerical statistical data, much akin to
the hazard rate and cumulative hazard analysis in life expectancy and fatigue studies.
For non-constant $\gamma(x)$ we hence have generalized the concept of cumulative hazard
and hazard rate.

This somewhat longer paper is not intended to be a review in the classical sense.
The emphasis was put on unfolding of a simple original idea, investigating
the consequences and application possibilities of unidirectional growth and resetting.
We did not have the occasion to exhaustively cite and review many important works
on theory or applications on complex systems with random dynamics.
Concerning our contributions, we did not limit ourselves to a simple repetition of earlier
published results. Several new ideas both in theoretical derivations and
real world applications are presented here in this new context.

Closing this paper we outline some possible further research directions on the main topic
treated here, based on unresolved questions. In the theory of investigating the
stability of stationary PDF-s and the possible generalized formulas for entropy
and entropic divergence further master equation classes have to be investigated more closely.
In particular local and nonlocal transition rates should be considered together without
a restriction to one direction (growth only) and without selecting out only a single
state ($n=0$) as a special one. Also the interpretation of negative $\gamma_n$
values remained incomplete. 
Finally, a much greater number of application on real world data
shall be undertaken in the future. 
In particular it would be instructive  to gather complementary knowledge
about the micro-transition rates ($\mu_n$, $\gamma_n$, $w_{nm}$) and 
not only to fit the PDF-s.

%\textcolor{Magenta}{
%\begin{itemize}
%\item Summary Table of Rates and Stationary Distributions
%\item Restrictions on Entropy Formula Generalizations
%\item Negative Rates and anti-H-theorems
%\end{itemize}
%}

%%%%%%%%%%%%%% ACKNOWLEDGEMENT %%%%%%%%%%

\section*{Acknowledgment}

T.~S.~B. thanks the University Babe\c{s}-Bolyai in Cluj for the UBB Star Fellowship.
Discussions with Andr\'as Telcs, Levente Varga, Hawong Jeong, G\'eza T\'oth and Mounir Afifi
are hereby acknowledged. We thank Constantino Tsallis for encouraging to present a review sized publication
about the achievements of our simple model.
This work has been partially supported by the Hungarian National Bureau for Research, Innovation and
Development (NKFIH) under project Nr. K 123815 and by
the Romanian Research Council UEFSCDI, PN-III-P4-PCE-2016-0363.

%%%%%%%%%%%%%% BIBLIOGRAPHY %%%%%%%%%%%%%%

\end{document}